\documentclass[prx,aps,twocolumn,floats,superscriptaddress,longbibliography]{revtex4-2}
\include{header}
\usepackage{soul}
\usepackage{manfnt}
\usepackage{xcolor}
\usepackage{multirow}
\usepackage{amsmath}
\usepackage{graphicx}
\graphicspath{{../Figs/}}
\usepackage{amssymb}
\usepackage{epsfig}
\usepackage[utf8]{inputenc}
\usepackage[T1]{fontenc}
\usepackage{newtxtext,newtxmath}
\usepackage{pdfrender}
\newcommand{\gtappr}{{{\lower4pt\hbox{$>$} } \atop \widetilde{ \ \ \ }}}
\newcommand{\ltappr}{{{\lower4pt\hbox{$<$} } \atop \widetilde{ \ \ \ }}}

\usepackage{simplewick}
\usepackage[colorlinks = true]{hyperref}
\usepackage{xcolor}
\definecolor{ginger}{rgb}{0.69, 0.4, 0.0}
\definecolor{brightpink}{rgb}{1.0, 0.0, 0.5}
\definecolor{fuchsia}{rgb}{1.0, 0.0, 1.0}
\definecolor{islamicgreen}{rgb}{0.0, 0.56, 0.0}
\definecolor{vividburgundy}{rgb}{0.62, 0.11, 0.21}
\definecolor{darkorchid}{rgb}{0.6, 0.2, 0.8}
\newcommand{\bx}{{\textbf{x}}}
\newcommand{\br}{{\textbf{r}}}
\newcommand{\bk}{{\textbf{k}}}
\newcommand{\bK}{{\textbf{K}}}
\newcommand{\bG}{{\textbf{G}}}

\newcommand{\bR}{{\textbf{R}}}
\newcommand{\dg}{^{\dagger }}
\newcommand{\pmat}[1]{\begin{pmatrix}#1\end{pmatrix}}

\newlength{\figwidth}
\newlength{\shift}
\newcommand{\Ustar}{U}
\newcommand{\Gstar}{(\gamma_0)^2}
\begin{document}

\title{Topological Mixed Valence Model for Twisted Bilayer Graphene 
}

\author{Liam L.H. Lau}
\affiliation{
Center for Materials Theory, Department of Physics and Astronomy,
Rutgers University, 136 Frelinghuysen Rd., Piscataway, NJ 08854-8019,
USA}
\author{Piers Coleman}
\affiliation{
Center for Materials Theory, Department of Physics and Astronomy,
Rutgers University, 136 Frelinghuysen Rd., Piscataway, NJ 08854-8019, USA}
\affiliation{Department of Physics, Royal Holloway, University
of London, Egham, Surrey TW20 0EX, UK.}

\begin{abstract}
	Song and Bernevig (SB) have recently proposed a topological
	heavy fermion description of
the physics of magic angle twisted bilayer graphene (MATBG), 
involving the hybridization of flat
band electrons with a relativistic conduction sea. Here we
explore the consequences of this model, 
seeking a synthesis of understanding drawn from heavy fermion
physics and MATBG experiments.  Our work identifies a key
discrepancy between measured and calculated onsite Coulomb
interactions, implicating renormalization effects that are not
contained in the current model.  With these considerations in mind, we
consider a SB model with a single, renormalized onsite interaction
between the f-electrons, containing a phenomenological heavy fermion
binding potential on the moir\'e AA-sites.  This feature allows the
simplified model to capture the periodic reset of the chemical
potential with filling and the observed stability of local moment
behavior. 
We argue that a two stage Kondo effect will develop in MATBG as 
a consequence of the relativistic conduction band: 
Kondo I occurs at high temperatures,  establishing a
coherent hybridization at the $\Gamma$ points 
 and a non-Fermi liquid of incoherent
fermions at the moir\' e K-points; at much low temperatures Kondo II
leads to a Fermi liquid in the flat band.  
Utilizing an auxiliary-rotor approach, we formulate
a mean-field treatment of MATBG that captures this physics, 
describing  the evolution of the
normal state across a full range of filling factors. 
By contrasting the relative time-scales of phonons and valence
fluctuations in bulk heavy fermion materials with that of MATBG we are led to
propose a valley-polaron origin to the Coulomb renormalization and the
heavy fermion binding potential identified from experiment. We also
discuss the possibility that the two-fluid, non-Fermi liquid physics
of the relativistic Kondo lattice is responsible for the strange metal
physics observed in MATBG.
\end{abstract}

\maketitle
\section{Introduction}
The discovery of magic angle 
twisted bilayer graphene (MATBG), developing flat bands
at ``magic angles''\cite{eandrei2010,BistritzerMacDonald2011,Cao18a,Cao18b, andrei_graphene_2020, andrei_marvels_2021}, has
opened a new avenue for the exploration of quantum materials.
At integral filling, novel spin and
valley polarized \cite{lian_twisted_2021, liu_quantum_2019, repellin_ferromagnetism_2020, wu_collective_2020, pixley_ferromagnetism_2019} Mott insulators develop, 
which on doping transform into 
strange metals \cite{cao_strange_2020, lyu_strange_2021, das_sarma_strange_2022, cha_strange_2021,ghawri_breakdown_2022, arora_superconductivity_2020, polshyn_large_2019, jaoui_quantum_2022, codecido_correlated_2019} and
superconductors \cite{young2019, Cao18b}  that have attracted intense
theoretical study \cite{ramires_2018,haule_2019,song21, shi_heavy_2022,bultinck_ground_2020, xie_nature_2020,
zhang_correlated_2020, balents_superconductivity_2020,
lian_twisted_2019, liu_nematic_2021, liu_chiral_2018, wu_theory_2018,
xie_topology-bounded_2020, xu_topological_2018,
lewandowski_pairing_2021, patrickfractional2020, khalaf_charged_2021, konig_spin_2020,
chichinadze_nematic_2020, gonzalez_kohn-luttinger_2019,
guinea_electrostatic_2018, huang_antiferromagnetically_2019,
isobe_unconventional_2018,  julku_superfluid_2020,
kennes_strong_2018,bascones_2023, eunahfractional2022, eunahfractional2023}. 
It is as if by 
tuning the gate voltage one can now explore a family of compounds
along an entire row of the periodic table. { This {\sl ``gate-tuned
chemistry''} poses a novel challenge to theoretical work.}

Various experiments suggest that electrons localized in the moir\' e
hexagons of MATBG resemble quantum dots \cite{cascade, diracrevivals, saito_isospin_2021}, forming 
localized  moments with valley and spin degeneracy near integer filling.
This evidence includes the lifting of spin/valley degeneracy 
observed in Landau fans,\cite{efetov1} 
a field-tunable 
excess electronic entropy at integer
filling \cite{weissmann} 
and the appearance of upper and lower Hubbard band-like features in
scanning tunneling microscopy measurements\cite{yazdani1}.
While the Bistritzer
MacDonald \cite{BistritzerMacDonald2011} model for magic angle graphene, 
provides an accurate description of the
plane-wave single-particle physics,
the presence of local moments, governed by short-range Coulomb interactions
underlines the importance of
developing  a real-space description of the physics, whilst taking the
topology of the system into account
\cite{ramires_2018,haule_2019,song21, shi_heavy_2022,song_all_2019,
zou_band_2018, efimkin_helical_2018, nuckolls_strongly_2020,
Pierce2021, hejazi_hybrid_2021, saito_hofstadter_2021, choi,
wu_chern_2021, kang_non-abelian_2020, po_faithful_2019,
lian_landau_2020,bascones_2023}. 

Various  works have suggested a close analogy between MATBG
and f-electron heavy fermion
materials\cite{ramires_2018,haule_2019,song21, shi_heavy_2022},
opening the problem up to the diverse conceptual and computational
methodologies developed for these systems. This paper explores the
implications of a recent theory 
by Song and Bernevig (SB)\cite{song21, Lau23} which
succinctly describes TBG as a topological {\sl
heavy fermion} problem 
: 
rather remarkably,
the moir\'e potential 
focuses the 
electron
waves into Wannier states that are tightly 
localized at the center of each moir\'e hexagon. 
These localized ``heavy fermions'' carry 
 spin ($\sigma=\pm 1 $) - valley ($\eta = \pm 1$) 
and orbital ($\alpha=\pm 1$) quantum numbers, forming 
an eight-fold degenerate multiplet that becomes mobile through the
effects of valence fluctuations into a topological conduction band. 

\begin{figure}[h!]
	\centerline{\includegraphics[width=\columnwidth]{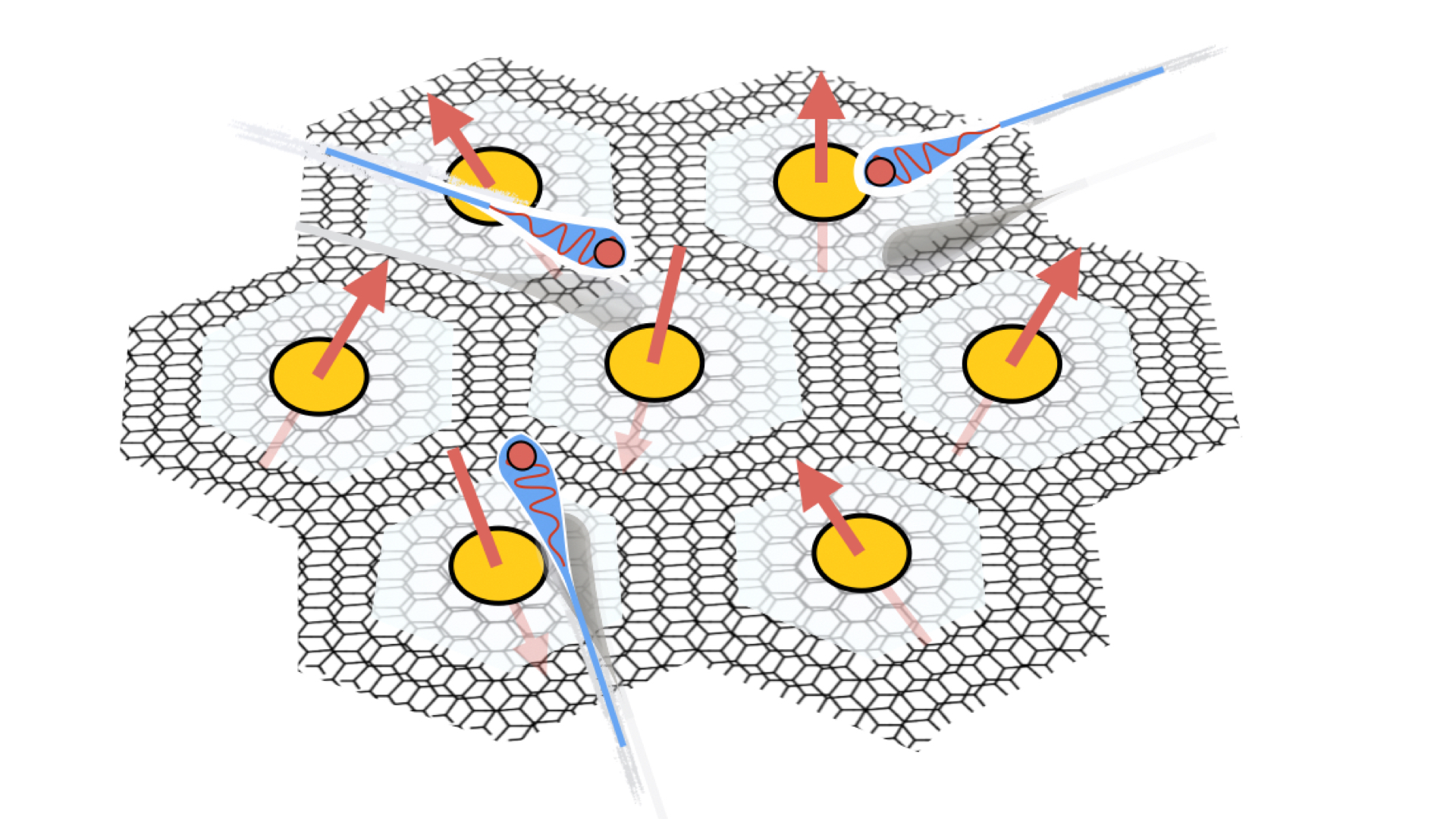}}\vskip -0.2cm
	\caption{Hexagonal lattice of exponentially localized Wannier $f$-states (orange) on each moir\' e AA site submerged in a sea of topological relativistic $c$-electrons (blue).}\label{fig:fig1}
\end{figure}

\begin{figure*}[tbh]\vspace*{-0cm}\centerline{\includegraphics[width=\figwidth]{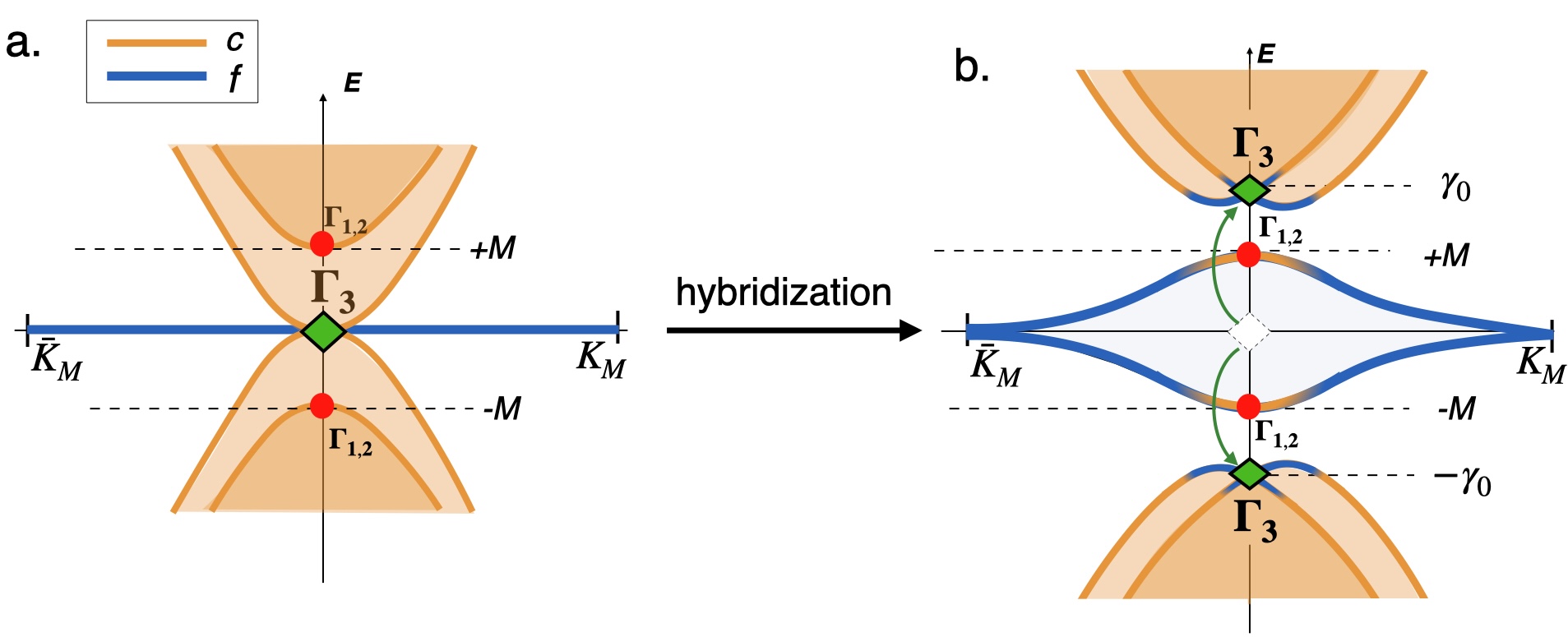}}\vskip
-0.2cm \caption{Schematic illustration of the Song-Bernevig model \cite{song21, Lau23},
showing {\bf a)} the unhybridized flat (f) -band
(blue) and relativistic  conduction band (c) (orange) with 
quadratic touching $\Gamma_{3}$ symmetry (green diamond) and $\Gamma_{1,2}$
symmetry points (red disk) with an excitation gap $M$; {\bf b)} hybridization causes band-inversion
between the $\Gamma_{3}$ and $\Gamma_{1,2}$ points, injecting topological
conduction states into the flat f-band.}\label{fig:fig2schema}\end{figure*}

The hybridization
of these flat-band (``f'') Wannier-states 
with 
a topological
conduction (``c'') band
captures the essential 
mirror, time reversal and particle-hole 
symmetries 
of the Bistritzer-Macdonald model\ (Fig. \ref{fig:fig1}).
In particular, the SB model establishes the correct band symmetries 
at the $\Gamma_{M}$ and $M_{M}$ points of the Brillouin zone,
giving rise to a pair of 
Dirac cones of the same chirality at the $K_{M}$ points of each valley. 
The $C_{2z} T$ and particle hole symmetry anomalies
responsible for the Dirac cones are reproduced by a
quadratic conduction band touching at $\Gamma_M$:
when hybridization is turned on, this anomaly is
injected into the f-electron band (Fig. \ref{fig:fig2schema}). 

Previous studies on the SB model\cite{sankarkondo,tsvelikkondo,tsveliksymmetric,
dzeromixedvalence,songkondo,matbgdmft} have adopted a fixed neutrality
assumption, in which the onsite Coulomb interaction takes the form
$\frac{U}{2} (n_{f}-4)^{2}$ 
and departures from neutrality are accomplished by varying
a uniform chemical potential, while treating all other interactions in
a Hartree-Fock approximation. 
 Various approaches have been used to describe the onsite physics,
including auxiliary bosons\cite{dzeromixedvalence}, impurity
approximations with Wilsonian renormalization group\cite{songkondo},
dynamical mean field theory \cite{haule_2019,matbgdmft} and a slave
boson approach \cite{dzeromixedvalence}. 

\figwidth=0.75\textwidth
In this paper, we build on these early Kondo lattice
models of MATBG, seeking to combine the key features of the SB model
with insights drawn from experiment and bulk heavy
fermion physics. 
In this paper, we focus on the paramagnetic phases of MATBG that
develop away from integral filling factors. 
 Highlighted aspects of our work are as follows:
\begin{itemize}
\item {\bf A discrepancy} identified between the ab-initio
scale of onsite interactions $U_{0}\sim 100${meV} in the SB model,
and the experimentally observed values of $U\sim 30$meV, pointing to
renormalization effects that lie beyond the current model.

\item {\bf An f-electron Binding potential}
of strength $-\Ustar\kappa \nu$, introduced phenomenologically to model 
the periodic reset in chemical potential and the 
diminished average inverse compressibility 
$\Delta \mu/\Delta \nu = \Ustar (1-\kappa)$. 



\item {\bf An auxiliary rotor 
}
mean field theory
\cite{slaverotor1, slaverotor2} to describe the alternating
patterns of Kondo and valence fluctuations in the paramagnetic phases across all filling
factors $\nu\in [-4,4]$. 

\item {\bf Two temperature scales.
}
The Dirac character of the conduction sea in MATBG profoundly affects the Kondo effect, leading to two competing
fixed points: a high temperature scale $T_{K}^{(1)}$ associated with
the formation of the topological band and a considerably lower
temperature $T_{K}^{(2)}$ associated with the development of coherence at the moir\' e K points. 
\end{itemize}

Key to the heavy fermion model of MATBG, is an
understanding of the binding potential that stabilizes localized
f-states at various integer filling factors $\nu$. 
{The simplest  view treats 
the back-gate of MATBG as a capacitor
that subjects the conduction
sea and f-level to a single chemical potential; in this picture,
a unit increase in the filling requires 
the chemical potential to shift by the Coulomb
energy, i.e  the coarse grained inverse compressibility is given by 
$\Delta \mu/\Delta \nu = U$.  Experimentally however, the
chemical potential rises by an
amount that is considerably smaller than the observed onsite
interaction.
For example, STM experiments at the AA-sites of MATBG observe cascades in the
electronic structure associated with an onsite $U\sim 30${meV} (blue read off from the upper and lower Hubbard band position from the AA data in \cite{cascade} and rounded up from $23 \pm 5 \, meV$), yet
the coarse-grained inverse compressibility is of
order $\Delta \mu/\Delta \nu \approx 15$meV\cite{cascade}. 
}

To shed further light on this physics it is useful to 
contrast  gate-tuned MATBG with bulk 
heavy fermion materials \cite{hewson, kondo_resistance_1964,
kondo_anomalous_1962, colemanmag, si_heavy_2010,
wirth_exploring_2016, stewart_heavy-fermion_1984,
dzero_topological_2010}, where the f-states are bound by a
nuclear potential whose depth progressively increases with atomic number
$Z$, accommodating the electron repulsion that rises with the filling
factor. The slow rise and periodic reset in the chemical potential
with filling factor  observed in MATBG suggests 
a corresponding binding mechanism 
in MATBG: 
We choose to incorporate the observed
physics in a phenomenological Hamiltonian 
\begin{equation}\label{Andersonlattice}
	H = H_{0} 
	+ \frac{\Ustar}{2}\sum_{\bR } (\hat \nu_{f{\bR }}-\kappa \nu)^{2},
\end{equation}
where $H_{0}$ is the non-interacting SB Hamiltonian, combined with 
an onsite Coulomb interaction of strength $\Ustar$ amongst the f-electrons
centered in the moir\'e unit cell at $\bR $, where $\hat  \nu_{f\bR }=\hat
n_{f\bR }-4$ is their number operator relative to half filling. 
This shifts the neutrality
point of the Coulomb interaction to $n_{f}=4+\nu$, 
where $\nu\propto V_{g}$ is directly proportional to the back-gate
voltage. 

While we introduce the attractive binding potential $-U \kappa \nu\; \hat{\nu}_{f \bR}$ phenomenologically to explain the partial chemical potential reset seen in experiment \cite{cascade, saito_isospin_2021, Pierce2021, diracrevivals}, this mechanism appears implicitly in previous DMFT+Hartree studies \cite{bascones_2023, matbgdmft} from an ad-hoc Hartree treatment of the interactions between the dispersive and localized states \cite{bascones_2023, song21, matbgdmft}. Our work explicitly highlights the vital role of this term in producing the chemical potential resets and provides a new interpretation of this term as an attractive binding potential for localized states. Given the uncertainties in the Hartree approximation, particularly when the onsite interaction is treated at much higher order in DMFT and the c-f hybridization is handled dynamically, we have adopted a phenomenological approach which leaves the microscopic origin of the binding potential to future debate.

There are two salient insights from our approach which we discuss in
depth at the end of this paper. Firstly, contrary to heavy fermion
materials. the optical phonon dynamics in twisted
bilayer graphene are fast compared to valence fluctuations of the
localized f-electrons. The phonons modify the ``atomic'' Hamiltonian
for the localized f-electrons in MATBG, giving rise to a Holstein
model of the form
\begin{equation}\label{eq:Holsteinsketch}
	H = \omega_{0}b\dg b +  g (b + b\dg )f\dg \tau_{x}f,
\end{equation}
where $b\dg$ creates the optic phonon and the $\tau_{x}$ couples to
the valley degree of freedom.  Bridging disparate fields, this leads
us revive a previously abandoned concept \cite{Sherington76, Riseborough87} from 
heavy fermion physics, proposing that the slow valence fluctuations in MATBG 
are dressed by inter-valley phonons, forming inter-valley polarons. 
Such polarons address the disparity between the measured and calculated onsite Coulomb by inducing the necessary renormalizations, but also could provide a natural contribution to the emergent heavy fermion binding potential $-\kappa U$. 
\begin{equation}
	\kappa U = \frac{g^2}{\omega_0},
\end{equation}
where $g$ is the electron-phonon coupling in
Eq. \ref{eq:Holsteinsketch}. This new perspective offers an attractive
synthesis of the high energy electron-phonon physics, notably present
in superconducting samples \cite{alietal2023}, and low energy strong
correlation physics in MATBG.

A second key insight is that the relativistic character of the Kondo
lattice in the SB model sets it apart from conventional bulk heavy fermion
physics. In the corresponding single impurity problem, a 
linear conduction density of states generates non-Fermi
liquid Kondo screening behavior, with a residual entropy 
\cite{Fritz_Vojta_2013}. We posit that
the topological character of the flat bands protects this physics in
the SB Kondo lattice, providing a possible origin for the observed strange
metal phenomena in these materials. 
We present a demonstration
that this framework yields two characteristic energy scales: a
high-temperature scale, due to scaling away from an unstable
Withoff-Fradkin fixed point \cite{withofffradkin}, corresponding to
the onset of topological effects, and a low-temperature scale
associated with the emergence of flat band coherence and scaling
towards a conventional heavy Fermi liquid.

We note that while strain effects \cite{Parker2021, Kwan2021, Wagner2022, yazdanitextures, jonahstrain2024}, and substrate alignment \cite{alignment1,alignment2}, particularly with hexagonal boron nitride (hBN), have been demonstrated to impact certain MATBG samples,  these effects are omitted here. 
A brief discussion of these effects is included at the end of the paper. 

The outline of the paper is as follows. Section \ref{sec:SBmodel}
reviews the
Song Bernevig model; in 
\ref{sec:interactingmodel} 
we use a renormalized Anderson model for MATBG to account for the behavior of both the chemical $\mu[\nu]$ and inverse compressibility $d\mu[\nu] / d \nu$  as functions of the filling $\nu$ within the ``atomic'' limit of the model.
Section \ref{sec:broadercon} discusses the effects of turning on
interactions in MATBG, identifying the two characteristic scales of
the underlying Landau Fermi liquid. 
Section \ref{whit} examines Kondo scaling in MATBG, using the single
impurity limit to gain insight into the lattice physics,  arguing that
the presence of a ``Withoff-Fradkin fixed point'' in the weak-coupling
physics introduces two 
characteristic temperature scales in MATBG.  In section
\ref{sec:slaverotor} we employ the auxiliary rotor mean-field
approach \cite{slaverotor1, slaverotor2} to describe the low energy
physics of our periodic Anderson model for MATBG, the strength of the
auxiliary rotor method is that it is exact in the strong and weak
coupling limits, hence we are able to capture the valence fluctuations
and Kondo effect in MATBG at all filling factors using a single
theory. \ref{sec:meanfieldresults} describes the results of
calculations using the rotor method. Finally, section \ref{sec:discussion}
discusses the physics beyond the current model,
including the origins of the renormalization of
the onsite Coulomb $\Ustar$ and the emergent heavy-fermion
potential $-\Ustar\kappa \nu$, speculating on the implications of our
findings for the future understanding of MATBG.

\section{Song Bernevig Model}\label{sec:SBmodel}

The one-particle Hamiltonian of the 
SB model 
\begin{equation}\label{}
H_0 = H_c + H_{fc} - \mu \hat N,
\end{equation}
hybridizes exponentially localized
Wannier $f$-electron states centered on the moir\' e AA sites with
topological conduction electrons defined by the 
Hamiltonian
\begin{eqnarray}\label{eq:thfconduction}
	H_c &=& \sum_
{\substack{|\textbf{k}|<\Lambda_c\\
	a,a' \eta\sigma  }}
c^{\dagger}_{\textbf{k} a\eta\sigma }\mathcal{H}_{aa'}^{\left(\eta\right)} \left(\textbf{k}\right) c_{\textbf{k} a'\eta\sigma  }.
\end{eqnarray}

Here  $c\dg_{\bk a\eta \sigma }$ creates a conduction electron with
orbital, valley and spin quantum numbers $a\in (1,4)$, $\nu=\pm$
and $\sigma =\pm 1$ respectively. The conduction electron dispersion 
\begin{eqnarray}\label{eqx:conductiondispersion}
	\scalebox{0.95}{$\mathcal{H}^{\left(\eta\right)}\left(\textbf{k}\right) = \begin{pmatrix}  & {\rm v}_{\star}\left(\eta k_x \alpha_0 + ik_y \alpha_z\right) \\
	{\rm v}_{\star}\left(\eta k_x \alpha_0 - ik_y \alpha_z\right) & M \alpha_x \end{pmatrix}$}
\end{eqnarray}
describes the momentum-dependent mixing between the four orbitals in each valley $\eta $. 
The off-diagonal terms give rise to 
an asymptotically  linear dispersion with velocity ${\rm v}_{\star }$, where 
the  Pauli matrices
$\alpha_{\mu}\equiv (\alpha_{0},\vec{\alpha})$ ($\mu=0,3$) act on 
the two dimensional blocks. 
The first two entries of the matrix ($a=1,2$) refer to electrons with
$\Gamma_{3}$ symmetry at the $\Gamma_{M}$,
point $\bk =0$, while the lower block-diagonal $a= (3,4)$
describes two orbitals of $\Gamma_{1}$
and $\Gamma_{2}$ symmetry, split by a mass $M$.  

$H_{c}$ gives rise to four bands with a 
four-fold spin-valley degeneracy at each $\bk $.
The low energy dispersion is quadratic at
$\Gamma_M$ and becomes relativistic
$|E|\sim {\rm v}_{\star} k$ at energies
$|E| \gtrsim M$, with a bandwidth $D \sim {\rm v}_{\star} K_{\theta}$.
The single-particle model for TBG in each valley has
a symmetry anomaly in the $C_{2z}T$ and particle-hole $P$ symmetries,
corresponding to two 
Dirac cones at the Fermi-level with 
the {\sl same} chirality.
Since the local orbitals are
topologically trivial, the unhybridized conduction electron
band-structure carries the symmetry anomaly.

The hybridization
between the conduction sea and $f$ electrons at each moir\' e AA site
$\bR$ is described by 

\begin{eqnarray}\label{eq:thfhybridization}
	H_{fc} &=&
	\gamma_{0} \sum_
{
	\textbf{R}\alpha \eta \sigma } 
\left( 
f^{\dagger}_{\textbf{R}\alpha \eta \sigma } 
	c_{\textbf{R} \alpha \eta \sigma  }  + \text{h.c.}\right).
\end{eqnarray}
Here
$f\dg_{\bR \alpha\eta }$ 
creates an f-electron with $\Gamma_{3}$ symmetry, orbital character $\alpha = 1,2$, valley
and spin quantum numbers $\eta $ and $\sigma $. The total degeneracy
of the bare f-states is thus $2 N_{f}=8$.
$c\dg_{\bR \alpha \eta}$ creates a c-electron in 
a non-exponentially
localized Wannier state 
centered at $\bR$ with the same $\Gamma_{3}$ symmetry, orbital, spin
and valley quantum
numbers as the f-electron; it
is related to the normalizable continuum $c\dg_{\bk a \eta \sigma}$
states by 
\begin{equation}\label{eq:conductionwannier}
	c_{\bR \alpha \eta \sigma} = \frac{1}{\sqrt{N_s}} \sum_{
\substack{ {\bk }, {\bf G}, a
\\
|\bk+{\bf G}| < \Lambda_c }} e^{i \bk \cdot \bR } [\phi^{(\eta)} (\bk+
{\bf G}, \gamma_{0})]_{\alpha a} c_{\bk+{\bf G} a \eta \sigma}.
\end{equation}
where the sum over all momenta has been divided up into a sum over reciprocal lattice vectors ${\bf G}$ of the
moir\' e lattice and a sum over momentum $\bk $ restricted to the
first moir\'e Brillouin zone. 
The matrix form factor is
\begin{eqnarray}\label{eq:hybridizationmatrix}
	\phi ^{\left(\eta\right)}\left(\textbf{k}\right) = e^{ \frac{-|\bk|^2 \lambda^2}{2} }\begin{pmatrix}
	 -\alpha_{0} + a_{\star}\left(\eta k_x \alpha_x + k_y \alpha_y\right),  & 0_{2\times2} \end{pmatrix},
\end{eqnarray}
where
$a_{\star}$ sets the length scale of the
hybridization and $\lambda$ is a damping factor proportional to the real
space spread of the localized f-Wannier states. Remarkably, the 
focusing effect of interference of the moir\'e potential produces
Wannier states of size $\lambda\sim a_{M}/5$, about a fifth of the moir\'e unit cell
size $a_{M}$. 
The natural bandwidth of the free theory is given by
$D \sim {\rm v}_{\star} K_{\theta}$, but after hybridization, $M$ becomes
the bandwidth of the moir\' e flat bands, while $\gamma_{0}$ is the energy of the $\Gamma_3$ irrep of the higher energy bands at the $\Gamma_M$ point.
In MATBG the case  where $M=0$ corresponds to the special {\sl chiral} limit of
twisted bilayer graphene, where the hybridized f-c band is 
{\sl completely} flat \cite{chiral1,chiral2}. 

Finally, $\hat N$ is the total electron
count measured relative to neutrality,
\begin{eqnarray}\label{}
\hat N & =&  \sum_
{\substack{|\textbf{k}|<\Lambda_c\\	a \eta\sigma  }}
(c\dg_{\bk a \eta\sigma}
c_{\bk a \eta\sigma}- \frac{1}{2}) + \sum_{\bR \alpha \eta \sigma
} (f\dg_{\bR  \alpha \eta \sigma }f_{\bR  \alpha \eta \sigma }-\frac{1}{2}),\cr &&
\end{eqnarray}
and $\mu$ is the chemical potential.

The approximate scales for the parameters in the SB model 
are 
\cite{song21, Lau23}
$D= v_{\star }K_{\theta }\approx 133$meV,
$\gamma_{0}=25$meV, $M=$3.7meV, $v_{\star }= -4.3$eV\AA, $K_{\theta
}=0.031 \hbox{\AA}^{-1}$, $a_{\star } = 65$\AA, and
$\lambda = 0.225 a_{M}= 29$\AA \ for the size of the Wannier states, 
which gives $\tilde{\lambda}=\lambda K_{\theta}= 0.90$.

The non-interacting SB model reproduces the band-structure of the
Bistritzer MacDonald model, giving rise to a central band of width
$2M$, split-off by an energy $\gamma_{0}$ from the upper and lower
bands as illustrated in Fig. \ref{fig:fig2schema}. The central band can contain up to $8$ electrons, and in the
non-interacting model, an applied
chemical potential causes the band-structure to move rigidly,
so that by changing the chemical potential $\mu$ over a range from
$-M$ to $M$, the electron count per moir\'e unit cell can be tuned from 0 to
8.

We note that the chiral limit \cite{chiral1,chiral2} of the Bistritizer MacDonald model, corresponding to the limit with no tunneling between graphene layers in the AA-stacked regions in TBG, is reproduced when $a_{\star} = 0$ \cite{song21, Lau23}. The central bands become exactly flat at the chiral limit magic angles \cite{chiral1}, corresponding to $M = 0$ in the SB model \cite{song21, Lau23}.

\section{Interactions}\label{sec:interactingmodel}

An appeal of the SB picture, is that it offers the possibility of a
simplified model for the interactions between the highly localized
f-electrons. Song
and Bernevig have provided a detailed calculation of the projection of
the Coulomb interaction, screened by the backgate, 
onto the Fock space of f- and c- electrons, which reveals a 
heirachy of interactions.   

In a conventional metal, electron charges are compensated by
the static charge of the background ions, but in MATBG, departures
from neutrality are compensated by screening charges in the back-gate
\cite{song21, Lau23}. In MATBG, Gauss's law enforces a
strict linear relationship between the electron filling of the flat bands
and the gate-voltage. 
For a filling $\nu$ the excess charge on the TBG
is  $q= eN_{M}\nu$ where $N_{M}$ is the number of moir\' e cells.
Gauss's law enforces  $q=CV_{g}$, where $C$ is the capacitance of the
backgate-dielectric-MATBG stack and $V_{g}$ the gate voltage.  Thus it
follows that the filling factor
\begin{eqnarray}\label{fillingcapacitor}
	\nu = \frac{q}{eN_{M}}= \frac{C V_{g}}{e N_M}  = \frac{V_{g}}{\Delta V_{g}}
\end{eqnarray}
is a strict linear function of the gate voltage. 
In essence then, the gate voltage {\sl is} the filling
factor, enforcing a canonical ensemble of definite filling factor
$\nu$ on the flat bands. 

The projected Coulomb interaction, screened by the backgate,
contains two major onsite interactions, an f-f Coulomb
interaction $U_{\text{screen}}$ and a f-c Coulomb repulsion $W$. 
The scale of the instantaneous Coulomb interactions is determined
by the 
Coulomb integral for an electron in one of the moir\' e Wannier
states, 
\begin{equation}\label{}
	U_{\text{screen}} = \int_{\bx,\bx '} \rho (\bx )V (\bx -\bx ')\rho (\bx ') 
\end{equation}
where 
\begin{equation}\label{}
\rho (\bx ) = \frac{e^{-x^{2}/\lambda^{2}}}{\pi\lambda^{2}}
\end{equation}
is normalized electron density in the Wannier state, while
\begin{equation}\label{}
V (\bx -\bx ') =
\frac{e^{2}}{4\pi\epsilon\epsilon_{0}}
 \sum_{n}
\frac{(-1)^n}
{\sqrt{|\bx -\bx '|^{2}+ (2 d n)^{2} }}
\end{equation}
is the Coulomb interaction between electrons, modified by the image charges.
Here $d$ is the distance to the back-gate. The summation is
$\sum_{n=0,1}$ for a single back-gate with a single image
charge per electron and $\sum_{n=-\infty}^\infty$
for a double back-gate, where there are 
the multiply reflected image charges of alternating sign behind the
back-gates. 
\figwidth=\columnwidth
\begin{figure}[tb]\vspace*{-0cm}\centerline{\includegraphics[width=\figwidth]{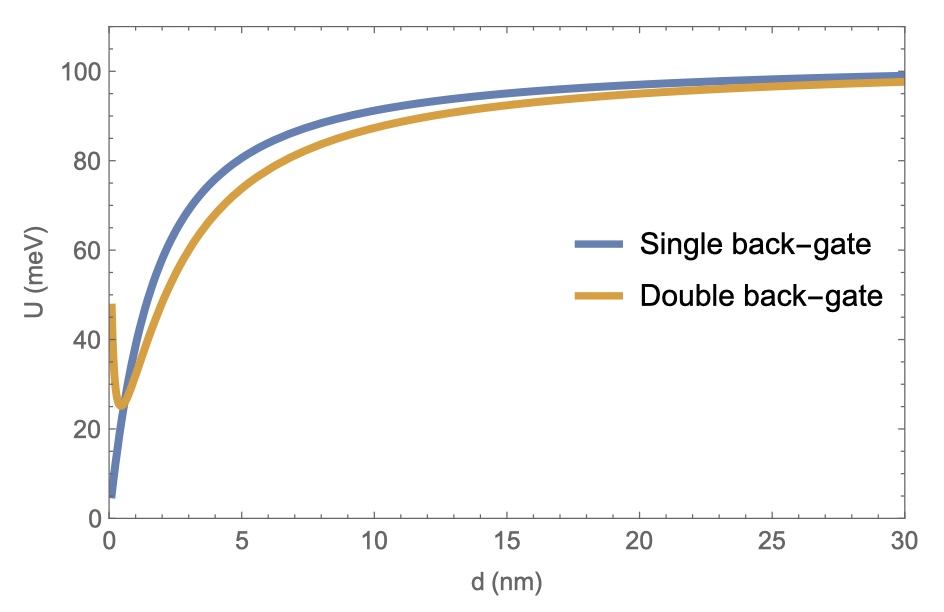}}\vskip
-0.2cm \caption{Showing the dependence of the Coulomb
integral as a function of back-gate distance for single and double
back-gate devices.}\label{uscale}\end{figure}

An evaluation of the Coulomb integral (Appendix \ref{AppendixA}) gives
\begin{equation}\label{screenyC}
	U_{\text{screen}} = 
U_{0}F\left[\frac{d}{\lambda} \right]
\end{equation}
where $F[x]$ is a screening function for the appropriate back-gate
(see Appendix \ref{AppendixA}) and
\begin{equation}\label{Coulomby}
U_{0}= 
\sqrt{\frac{\pi}{2}}\frac{e^{2}}{4\pi\epsilon\epsilon_{0}\lambda} \sim
103 \hbox{meV}
\end{equation}
is the unscreened Coulomb energy, estimated, following SB with 
$\epsilon=6.0$ as in-plane
dielectric constant for a Boron-Nitride substrate with $\lambda =
2.9nm$\cite{Laturia2018}. 
  Fig \ref{uscale} shows
the dependence of the Coulomb integral on the distance $d$ to the
back-gate, showing how the thinner the back-gate layer, the more
screened the Coulomb interaction. 
Song and Bernevig assume a distance $\xi = 2d\approx
10nm $ to the image charges in the 
back-gate.  With these values, \eqref{screenyC} predicts 
$U_{\text{screen}}\sim 70$meV, comparable with the value $U_{SB}\sim 58$meV 
obtained in SB.
This onsite interaction substantially exceeds the band-width 
of the flat band, leading to a situation in which the electrons are on
the brink of localization into states with integer occupations.  

There is however, an important discrepancy between the ab-initio
Coulomb interaction values calculated by Song and Bernevig, and that observed
experimentally. The STM experiments which reveal 
localized states at the AA moir\' e sites
\cite{cascade} indicate   
a Coulomb interaction 
$\Ustar\approx 30$meV (read off from the upper and lower Hubbard band position from the AA data in \cite{cascade} and rounded up from $23 \pm 5 \, meV$). 
In  fact the discrepancy with theory is even more substantial
when we take into account that 
the devices on which these measurements were made involve
a distance of about  $d=320 \, nm$ (combined thickness of around $40\, nm$ for the hBN and $285 \, nm$ for the $SiO_2$ \cite{privatecomms}) between the semiconductor back-gate and
the MATBG, for which we would expect the much larger value, $U\sim U_{0}$.
There is thus a factor of four discrepancy between the theoretical value
of $U$ and that observed in the STM cascade experiments. Indeed, the DMFT study \cite{matbgdmft} uses interaction scales \cite{song21,Lau23} calculated from back-gate distances which are more than an order of magnitude smaller than in the STM cascade experiments \cite{cascade}.

DMFT studies
have dealt with this discrepancy either by taking small back-gate distances \cite{matbgdmft} or by introducing a
large, phenomenological
dielectric constant $\epsilon=10-12$\cite{haule_2019,bascones_2023}  for the
substrate. This important
discrepancy between the measured and ab-initio Coulomb
interactions
hints at renormalization effects that lie beyond a static Coulomb
interaction. 

We note that previous DMFT works \cite{bascones_2023, matbgdmft} predict a Hubbard gap smaller than the Hubbard $U$ parameter. However, we believe that the renormalization factor between the gap and $U$ in these references (at best a factor of 2) is insufficient to reconcile the discrepancy between the experimentally measured gap with interaction scale $U = 23 \pm 5 \, \textrm{meV}$ \cite{cascade} and the much larger interaction scale $U_{\text{screen}} \sim 103 \, \textrm{meV}$ \eqref{screenyC} expected from the SB model using the experimental gate distance ($d = 320 \, \textrm{nm}$) from the Cascades study \cite{cascade}.

One possibility of renormalization effects beyond a static Coulomb interaction is dynamic screening by
the lattice, a point we return to in the
discussion.  
Here though, we assume that these effects can be taken account
by a renormalized Anderson model, in which the scale of the
interactions is set by experiment. 

\figwidth=\columnwidth
\begin{figure}[tb]\vspace*{-0cm}\centerline{\includegraphics[width=\figwidth]{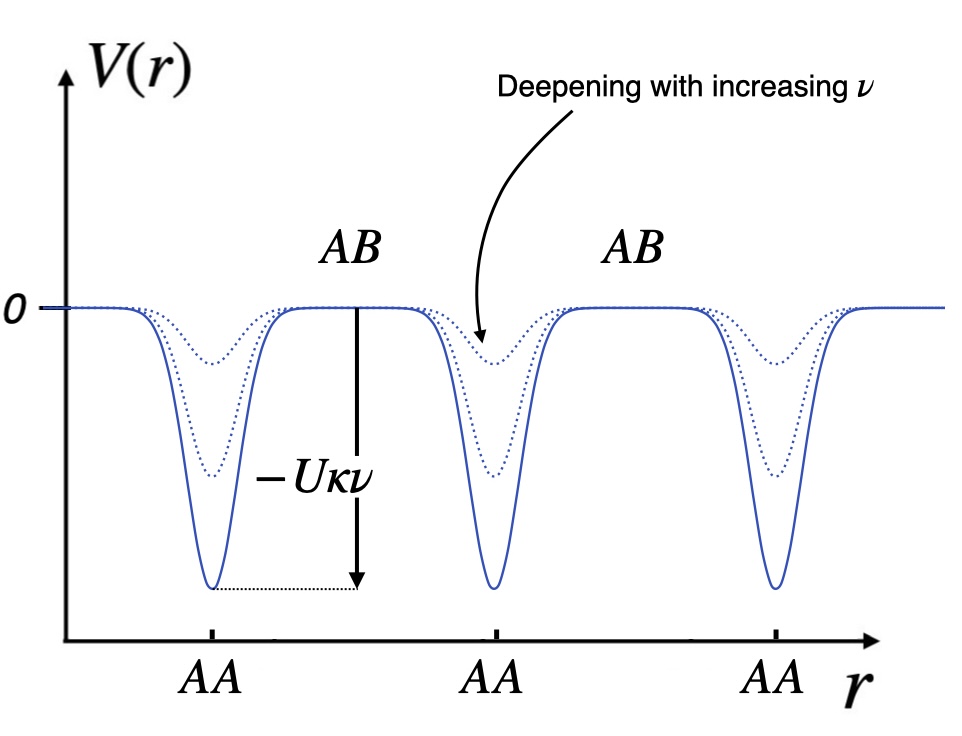}}\vskip
-0.2cm \caption{Emergent heavy-fermion potential on the AA-sites with
a depth proportional to the filling $\nu$. The well becomes 
progressively deeper with filling, offsetting the Coulomb
repulsion between f-electrons.}\label{hfpotential}\end{figure}

\begin{figure}[b]\vspace*{-0cm}\centerline{\includegraphics[width=\figwidth]{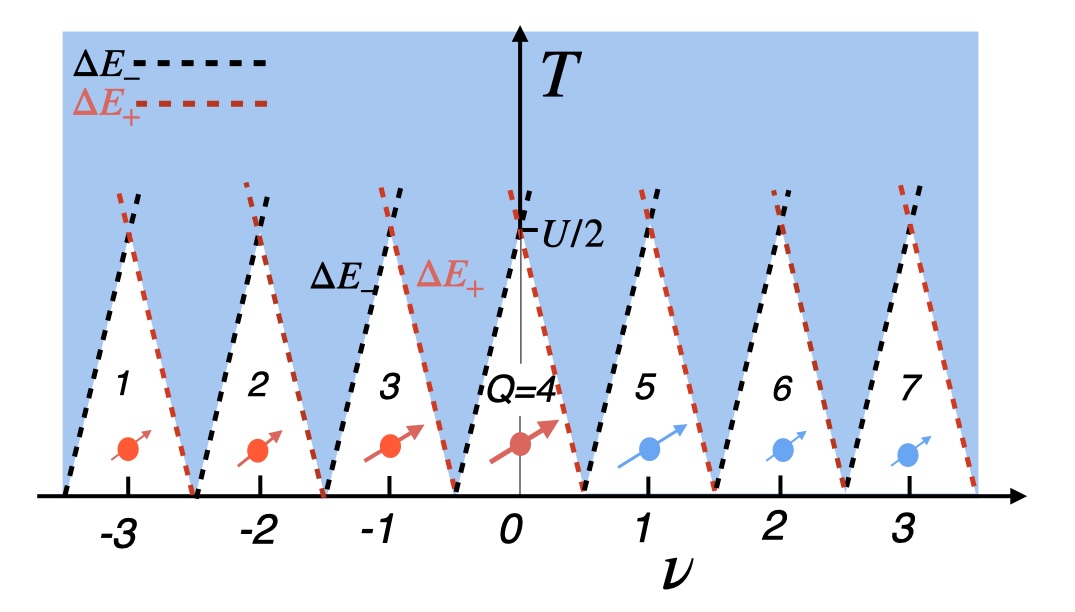}}\vskip
-0.2cm \caption{Sawtooth 
phase diagram for the ``atomic limit'' of MATBG, 
as a function of filling $\nu$.
White regions denote a stable local moment with 
$Q$ f-electrons, bounded by the ionization energies $\Delta E_{\pm}$
for adding or removing one electron. }\label{sawtooth}\end{figure}

\figwidth= 1.7\columnwidth
\begin{figure*}[tbh]\vspace*{-0cm}\centerline{\includegraphics[width=\figwidth]{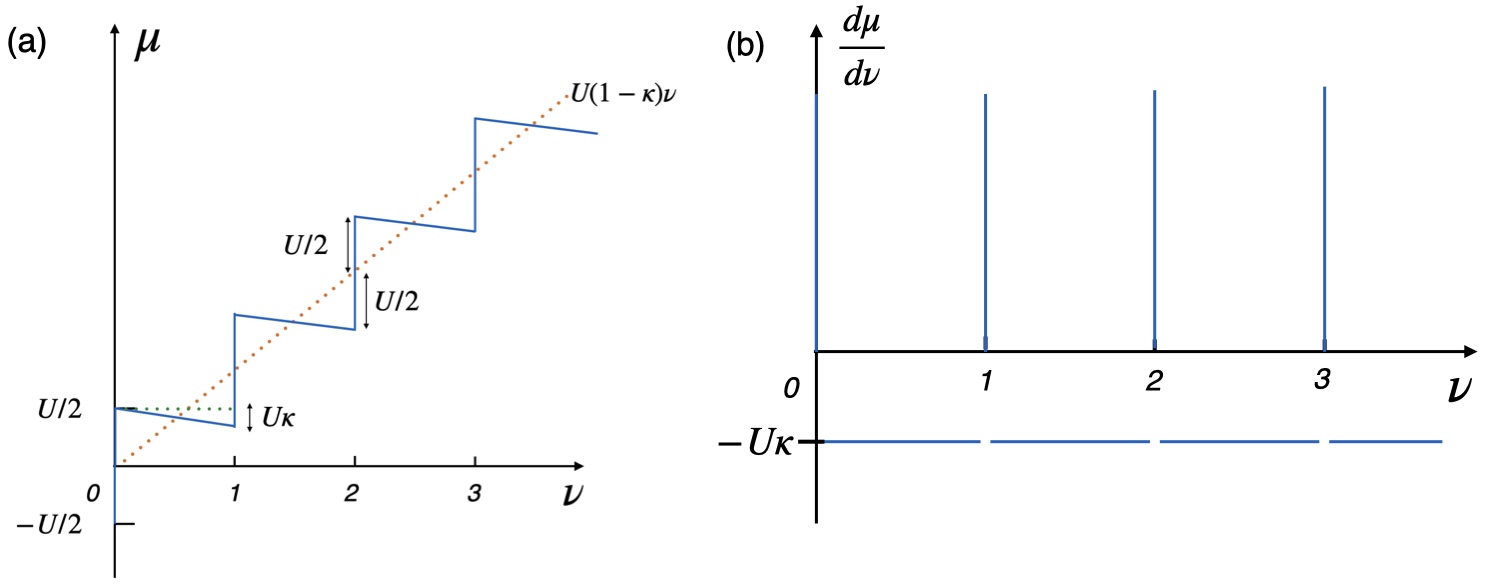}}\vskip
-0.2cm \caption{Sketch in blue of (a) the chemical potential $\mu$ and (b) the inverse compressibility $d \mu / d \nu$ as functions of filling factor $\nu$ for the ``atomic'' limit of the renormalized Anderson model for MATBG at zero temperature for nonzero $0 <\kappa <1$ .}\label{phenom}\end{figure*}

\subsection{Phenomenological Interaction 
Model}\label{phenomenologicalmodel}

Rather than pursuing a comprehensive microscopic approach, we adopt a
phenomenological strategy to capture the dependence of the 
interactions on filling factor. We
focus solely on the residual onsite interactions $\Ustar$ between
the f-electrons responsible for the observed  Coulomb blockade
physics. 
A second element of our treatment, is the introduction of an emergent heavy-fermion potential $-\Ustar \kappa \nu$ 
on the AA-sites (Fig. \ref{hfpotential}), which linearly deepens with backgate voltage $\nu
\propto V_g$ 
The renormalized Anderson model that we work with is the following,
\begin{eqnarray}\label{PhenomModel}
	H &=& H_{0} + \frac{\Ustar}{2}\sum_{\bR}  (\hat n_{f{\bR }}-4)^{2} - \Ustar\kappa \nu \sum_{\bR} \hat n_{f{\bR}}  \cr &\equiv & H_{0} 
	+ \frac{\Ustar}{2}\sum_{\bR } (\hat \nu_{f{\bR }}-\kappa \nu)^{2}.
\end{eqnarray}
The parameter $\kappa$ controls the emergent heavy-fermion
potential on the AA-sites, and we show in the following subsection
that a finite $\kappa$ is needed to understand the observed 
reset in the chemical potential $\mu$ that occurs as a function of
filling factor.

\subsection{Coulomb Blockade Physics}\label{coulombblockade}
\figwidth=\columnwidth

We begin by considering the unhybridized ``atomic'' limit 
of the renormalized Anderson model \eqref{PhenomModel}, given simply by 
\begin{eqnarray}\label{phenomatom}
	H_{A} (\bR ) = \frac{\Ustar}{2} (\hat  n_{f\bR }-4 -\kappa \nu)^{2} - \mu \hat n_{f\bR},
\end{eqnarray}
The physics here
is similar to a quantum dot. 
 The stability of the quantum dot with $n_{f}=Q$ f-electrons requires that the 
ionization energies (Appendix \ref{sec:phemchempot})
\begin{eqnarray}\label{eq:coulombblockade}
	\Delta E^{Q}_{\pm }=  E_{Q\pm 1}- E_{Q}  =  \frac{\Ustar}{2}\pm (\Ustar(1 - \kappa) \nu - \mu),
\end{eqnarray}
are both positive. Here we have set  $\nu = Q-4$.
The energies  $\pm\Delta E^{Q}_{\pm}$ describe the
offset location for the upper and lower Hubbard peaks in the
f-spectral function (Fig. \ref{FigEL}). At zero temperature, the local moment with $Q$ f-electrons is stable provided that the chemical potential satisfies
\begin{equation}\label{muconstraint}
	\Ustar/2 > |\Ustar(1- \kappa) \nu - \mu[\nu] |.
\end{equation}
To increase the filling factor by one, the chemical potential $\mu$ must jump by $\Ustar$ at each integer $\nu$. When continuing to fill the ``atomic'' model \eqref{phenomatom} from integer filling $\nu \rightarrow \nu +1$, the extra onsite Coulombic cost $\Ustar$ is partially offset by the emergent heavy fermion potential $\Delta E_f = -\Ustar\kappa$. Consequently the chemical potential must shift $\Delta \mu = \Ustar (1 - \kappa)$ to fully compensate the Coulombic cost (details in Appendix \ref{sec:phemchempot}).

At a finite temperature, 
the local moment with $Q$ f-electrons is stable provided 
\begin{equation}\label{}
	k_{B}T < \Ustar/2 - | \Ustar(1- \kappa) \nu - \mu[\nu]|
\end{equation}
which defines the saw-tooth phase diagram shown in Fig. \ref{sawtooth}.

In Fig. \ref{phenom} we illustrate the chemical potential $\mu$ and inverse compressibility $d \mu / d \nu$ as functions of the filling factor $\nu$, depicted in blue, for the ``atomic'' limit of the renormalized Anderson model \eqref{phenomatom} at zero temperature with a finite value of $\kappa$.

The presence of a finite hybridization will cause the f-valence 
to fluctuate through the virtual emission or absorption of electrons, 
 $f^{Q}\rightleftharpoons f^{Q-1}+e^{-}$ and
$e^{-}+f^{Q}\rightleftharpoons f^{Q+1}$. 
At energy scales below $\Ustar/2$, 
the physics of the low-energy region are then described
by a voltage-tuned ``Kondo lattice''\cite{swolf,coqblin}. 

The presence of a finite negative gradient in the chemical potential
$\mu[\nu]$ in between integer filling factors, coupled with evidence
of negative inverse compressibility $d\mu/ d\nu$ empirically suggests
a finite value of $\kappa$. 
From
\cite{cascade, saito_isospin_2021, Pierce2021, diracrevivals}, we
extract phenomenological values of $\Ustar \approx 30 \, meV$ and $\kappa \approx
0.8 $. We find that the bare Song-Bernevig hybridization $\gamma_{0}$ results in a significant smoothing of the sharp features in the chemical potential $\mu[\nu]$ as a function of filling found in the ``atomic'' limit. To preserve the partial resetting behavior, it is crucial to maintain the established hierarchy of energy scales within the heavy-fermion analogy. Specifically, the hybridization between the f-electrons and c-electrons must undergo renormalization to ensure that the reduced hybridization strength $\gamma_0 \ll U$. We defer the discussion of the possible mechanisms underlying
these phenomenological parameters and renormalizations to section
\ref{sec:discussion}.

\figwidth=\columnwidth

\begin{widetext}

\section{Qualitative Considerations}\label{sec:broadercon}

\subsection{Adiabatic Considerations: the Heavy Fermi Liquid}\label{}

The non-interacting SB model describes a narrow band of
f-electrons, with a linear Dirac dispersion of fixed chirality
centered the $K_{M}$ points with a Dirac velocity $v_{D} $ (see
Appendix \ref{sec:appendix}). 
\begin{equation}\label{diracv}
	{\rm v}_D \approx 
	3 \left(\frac{\gamma_K}{D}\right)^2 \frac{M}{D} \left(
	2(1+a_{\star}^2 K_{\theta}^2) + \tilde \lambda^2
(1-a_{\star}^2 K_{\theta}^2)\right){\rm v}_{\star } 
\end{equation}
\end{widetext}
where $D={\rm v}_{*}K_{\theta }$ and ${\gamma}_{K}= \gamma_{0}
e^{-{\tilde\lambda}^2/2}$ is the strength of the hybridization at the
$K_{M}$ point, where $\tilde{\lambda}= K_{\theta}\lambda$. The approximate
band-width of this flat band is given by $W=v_{D}K_{\theta }$ or
\begin{equation}\label{bwidthy}
	W \approx 
	3 \left(\frac{\gamma_K}{D}\right)^2 {M}\left(
	2(1+a_{\star}^2 K_{\theta}^2) + \tilde \lambda^2
(1-a_{\star}^2 K_{\theta}^2)\right).
\end{equation}
In the third chiral limit (flat limit) \cite{song21, Lau23}, where $M=0$, the flat band-width identically
vanishes.  Even in the non-interacting model, there are in fact two
important energy or temperature scales, a high temperature scale
${ T^{(1)}\sim \gamma_{0}\sim  25}$meV,  below which the excitations are confined within
the low energy flat band, and a much lower scale
\begin{equation}\label{}
	T^{(2)}\sim \left(\frac{\gamma_0}{D}\right)^2 { \left(2(1+a_{\star}^2 K_{\theta}^2) + \tilde \lambda^2
	(1-a_{\star}^2 K_{\theta}^2)\right) {M} \sim 1.0 \hbox{meV},}
\end{equation}
corresponding to the Fermi temperature of the flat band. 
This large separation of scales is a key feature of the SB model that
we expect to continue when interaction effects are taken into
account.

\begin{figure}[h]
	\centering
	\includegraphics[width = 0.8\columnwidth]{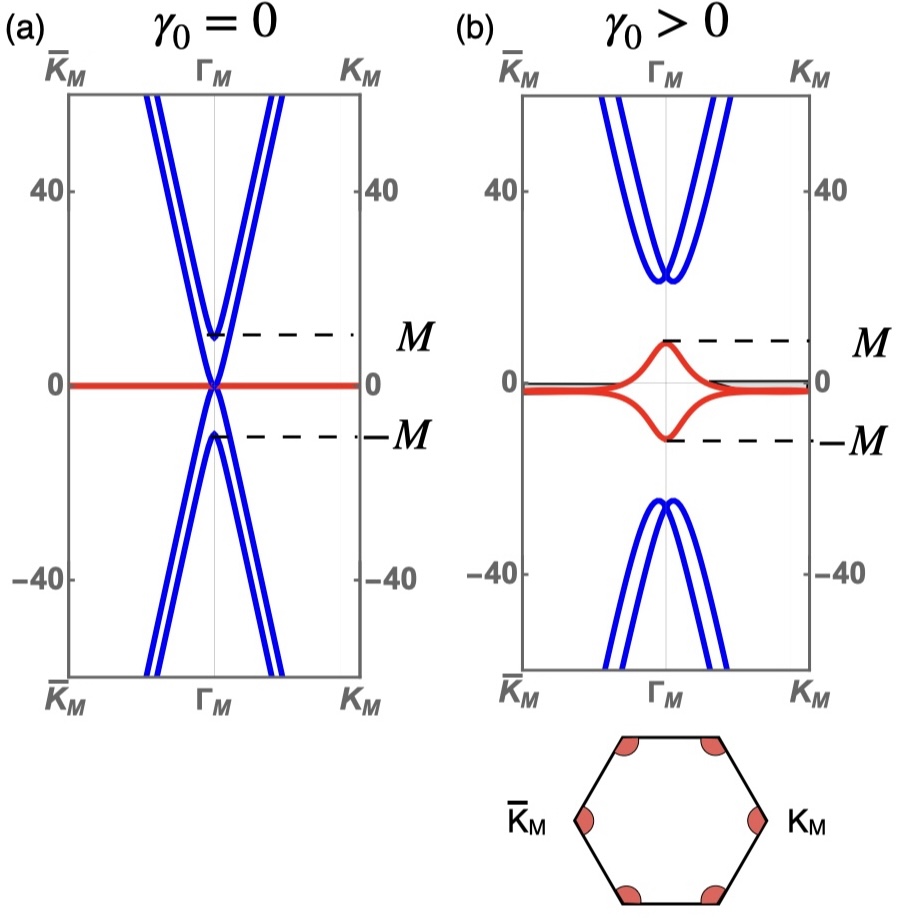}
	\caption{Schematic contrasting the non-interacting 
bandstructure of the SB model at fixed filling for a) zero and b)
	finite hybridization, with zero interaction. The Dirac points sink into the Fermi sea, producing two approximately circular Fermi surfaces of predominantly f-character, with four-fold valley spin symmetry, centered at each $K_{M}$ and $\bar{K}_{M}$ points. }
	\label{schematicadiabatic}
\end{figure}

\begin{figure}[b]\vspace*{-0cm}\centerline{\includegraphics[width=\figwidth]{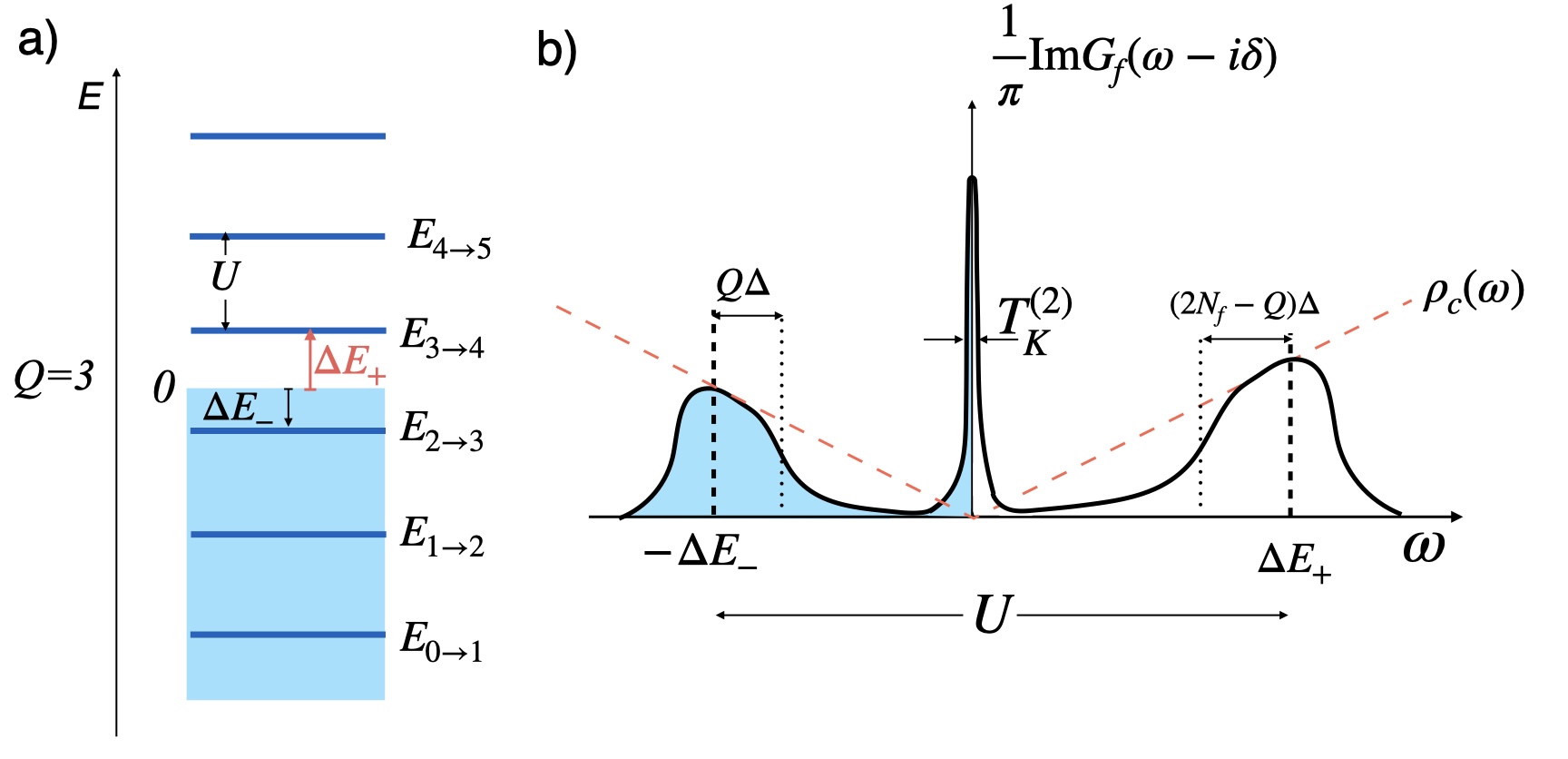}}\vskip
-0.2cm \caption{a) Energy level diagram
showing the position 
 f-level excitation energies 
 for the case of $Q=3$ f-elecrons, i.e 
 $\nu_0=-1$. b) Spectral function for the f-state in an impurity
 model, showing upper and lower Hubbard resonances and the central
 Kondo resonance against the background of the linear density of
 states of the conduction sea.  }\label{FigEL}\end{figure}

Provided $M>0$, a Fermi liquid forms and  on doping away from neutrality 
the Dirac points sink into the Fermi sea, producing two approximately
circular Fermi surfaces of predominantly f-character (Fig. \ref{schematicadiabatic}), with four-fold
valley spin symmetry,  centered at each $K_{M}$ and $\bar{K}_{M}$ points, each of area  $A_{FS}\sim  \pi k_{F}^{2}$
area which satisfies Luttinger's sum rule, which we can write as 
\begin{equation}\label{}
8\frac{A_{FS}}{A_{M}} = \nu
\end{equation}
where ${ A}_{M}$ is the area of the moir\'e Brillouin zone.
The non-interacting
f-electrons thus form a Dirac sea of relativistic chiral fermions with
a bandwidth of approximately $v_{D}k_{F}$, occupying a fraction $\nu/8$
of the Brillouin zone. 
The SB model also predicts that at the $\Gamma_{M}$ point, the 
energy eigenvalues are  $\epsilon_{\Gamma}=\left\{\pm M - \mu, \pm \gamma_{0} - \mu
\right\}$, where those with energy $\pm M$ are entirely of conduction
character, whereas those with energy $\pm \gamma_{0}-\mu$ are an equal
admixture of f and topological conduction electrons.  

Let us consider what happens when interactions are adiabatically
introduced at constant filling factor $\nu$ to produce a Landau Fermi
liquid: this requires that $|M|>0$.
Now the f-states will 
renormalize with a Quasi-particle weight $Z_{f}$ characterizing the
$K_{M}$ points of the Brillouin zone. 
So long as the
ground-state remains a Fermi liquid, the 
Fermi surface area remains an adiabatic invariant, which will cause
the f-states to remain pinned close to the Fermi energy, with energies
$\epsilon_{\bk } = \lambda \pm v^{*}_{D}|\bk - {\bf K_{M}}|$, here
$v^{*}=Z_{f}v_{D}$ is a renormalized Fermi velocity while 
in $\lambda \sim W^{*}=Z_{f}W$ is of order the renormalized
band-width.

\figwidth=1.1\columnwidth
\figwidth=\columnwidth

The principle energy scales of the SB Anderson lattice can be obtained
by considering a corresponding impurity Anderson model, formed from a single
moir\'e f-state embedded in a relativistic electron gas. 
The relativistic character of the
conduction sea gives rise to a density of states per moir\'e per
valley per spin,  that is linear in
energy at high energies. The density of states per spin per valley per
orbital in the $\Gamma_{3}$ channel that hybridizes with the f-states is
\begin{equation}\label{dos}
\rho_{c} (E)= \frac{A}{D^{2}}\times 
\left\{
\begin{array}{cc}
|E|
,& |E|>M\cr\cr
\frac{1}{2} (|E|+M),& |E|<M.
\end{array} \right.
\end{equation}
where $A = 2 \pi / (3 \sqrt{3})\approx 1.2$ (see Appendix \ref{sec:gamma3cdos}). 
In the presence of a chemical
potential, this density of states shifts downwards
in energy by an amount $\mu$, and now $\rho_{c} (E,\mu)= \rho_{c}
(E+\mu)$.

If we ignore the effects of interaction, the hybridization width
(half width at half maximum) of an
isolated non-interacting Anderson impurity using the bare SB hybridization $\gamma_{0} = 25 meV$ \cite{song21, Lau23} is given 
by 
\begin{eqnarray}\label{momint}
	\Delta_{0}[\mu]  &=& \pi\overline{\gamma^{2}_0(k)}\rho_c(\mu) 
\end{eqnarray}
where  $\overline{\gamma_0^2(k)} \approx 2 \gamma_0^2 $ is the momentum integrated average of the hybridization squared over a circle of radius $K_{\theta}$ (see Appendix \ref{sec:avghyb}). From \eqref{dos}, we obtain the noninteracting hybridization width of a single Anderson impurity at neutrality to be,
\begin{eqnarray}\label{noninthyb}
	\Delta_0 (\mu=0) = \frac{\pi A}{2} \frac{\overline{\gamma^{2}_0(k)}}{D^2} M \approx  0.1 M.
\end{eqnarray}
While, for the full lattice, the bandwidth of the SB model is $W = v_{D} K_{\theta}$. From \eqref{diracv}, we obtain
\begin{equation}\label{barebandwidth}
W = v_{D}K_{\theta }\approx  0.4 M.
\end{equation}
The two quantities $\Delta_0(\mu = 0)$ and $W$ are of comparable
magnitude at neutrality.

When the interactions are turned on, the f-spectral function splits
into an upper and lower Hubbard peak at locations $E^{Q}_{+}$ and
$-E^{Q}_{-}$, with a Kondo resonance in the center as shown in
Fig. \ref{FigEL}. The upper and lower resonances have a half-width of order
$(2 N_{f}-Q)\Delta_0 $ and $Q\Delta_0 $, respectively,  where $2 N_{f}=8$ for
TBG. 


\subsection{
Withoff-Fradkin Scaling  and Non-Fermi liquid physics}\label{whit}

Once the temperature drops below the characteristic energies for
valence fluctuations $T\ll{\rm min} (\Delta E^{Q}_{\pm})\sim \Ustar$,
we can integrate out the charge fluctuations to produce a low-energy
Kondo lattice description. 
The resulting low energy 
effective Hamiltonian, obtained by performing a Schrieffer-Wolff transformation \cite{coqblin} on \eqref{Andersonlattice}
\begin{equation}\label{eq:kondolattice}
	 H_{K} =  
	\sum_{\scriptsize{{\rm v}_{\star}|\bk| <D}}
 c\dg_{\bk} \mathcal{H}\left(\bk\right) c_{\bk} 
	+J_{\text{eff}}\sum_{\textbf{R} B
	B'} c^{\dagger}_{\textbf{R} B}
	c_{\textbf{R}B'} S_{B
	B'}\left(\textbf{R}\right)
\end{equation}
is a topological Kondo lattice model. 
Here
$B \equiv \left(\alpha \eta \sigma \right)_{2}\equiv  1,2\ldots 8 $ is
the $SU(8)$ index written in binary and 
\begin{equation}\label{}
	S_{BB'} (\bR ) = f\dg_{\bR  B}f_{\bR  B'} - \frac{Q}{2N_f}\delta_{BB'}
\end{equation}
is the SU (8) spin operator. The operator $c^{\dagger}_{\textbf{R} B} \equiv c^{\dagger}_{\bR \left(\alpha \eta \sigma\right)}$ defined in Eq. \ref{eq:conductionwannier}
 creates a spatially extended $\Gamma_3$ conduction electron state,
centered (rather than localized) at
$\bR$ with quantum numbers $B$. In the Kondo limit $\mu \ll \Ustar$, the strength of the effective Kondo interaction 
\begin{equation}\label{Jeff}
	J_{\text{eff}} = \sum_{\pm}
	\frac{\Gstar}{\Delta E_{Q\rightarrow Q\pm 1}}
	= \frac{ 4 \Gstar }{\Ustar}  \ F\left[\delta \nu\right]
\end{equation}
where $\delta \nu = \nu - \nu_0$ is the difference between the actual
filling $\nu$ and the integer ``atomic'' filling $\nu_0$, and
$F[\delta \nu]= 1/ (1- (2 \delta \nu)^{2})$.

\begin{figure}[t] \centering \includegraphics[width =
\columnwidth]{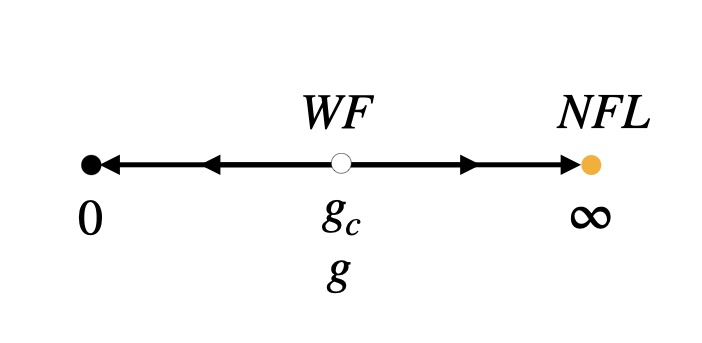} \caption{
	$M = 0$ scaling diagram for the dimensionless Kondo
	coupling $g$, showing the unstable Withoff-Fradkin (WF) fixed
	point $g_c$. Flow to weak coupling occurs below WF ($g< g_c$);
	above WF ($g>g_c$), flow to strong coupling yields a non-Fermi
	liquid (NFL) fixed point with residual entropy. 
}
\label{scalingschem1}
\end{figure}

To get an idea of the underlying physics of  the MATBG Kondo lattice, 
it is instructive to
consider the properties of a corresponding single-impurity model: i.e,
consider a thought experiment in which only one moir\' e AA site
is occupied with f-electrons. It is particularly instructive to
consider the symmetric neutral case where  $\nu=0$ and $M=0$
(the third chiral limit (flat limit) of MATBG \cite{song21, Lau23}). 
In this case $\rho_{c} (\omega)\propto |\omega|$.  
The Kondo coupling $J(\Lambda)$ is governed by the leading order scaling equation 
\begin{equation}\label{Jscalingregimes}
	\frac{\partial J }{\partial \ln \Lambda} = -2N_f J(\Lambda)^2
	\rho_c(\Lambda)+ O (J^{3}),
\end{equation} 
where
$N_{f}=4$ is the valley-spin degeneracy. Rewriting
\eqref{Jscalingregimes} as a dimensionless coupling constant $g(\Lambda)
\equiv J(\Lambda)\rho_c(\Lambda)$, we find that
the Kondo coupling constant renormalizes
according to the ``Withoff-Fradkin'' scaling equation\cite{withofffradkin}, 
\begin{equation}\label{scalingregimes}
\frac{\partial g }{\partial \ln \Lambda} = 
g (\Lambda) - 2 N_f g (\Lambda)^{2} + O (g^{3}),\qquad 
\end{equation}
where the first term derives from the derivative of the density of
states.
The competition between the linear and the quadratic terms in this
scaling equation gives rise to the unstable Withoff-Fradkin fixed
point (Fig. \ref{scalingschem1}), located in this case at $g_{c}= 1/
(2N_{f})=1/8$. Provided $g>g_{c}$, a Kondo effect does take place, characterized by a single Kondo temperature $T_{K}^{(1)}$. 
However, although $g$ now scales to strong coupling,  
the fully screened state that develops in a
Dirac sea is not a Fermi liquid, forming a Kondo
resonance with a singular density of states 
and a finite residual entanglement entropy\cite{Fritz_Vojta_2013}.  

\begin{figure}[b] \centering \includegraphics[width =
\columnwidth]{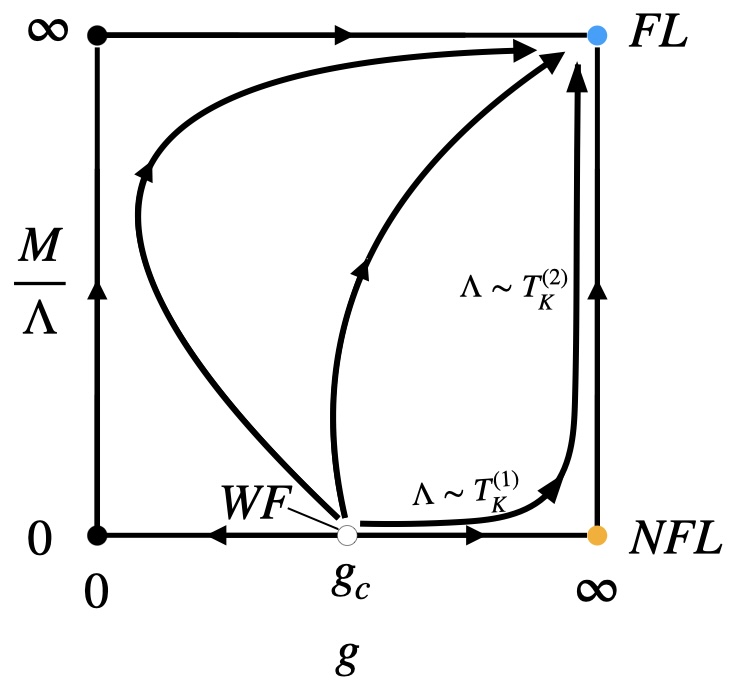} \caption{
Schematic scaling trajectories for the (single-ion) Kondo effect
within the SB model.  Finite $M$ ensures the Kondo effect for all $g$. For $g>g_c$ and small $M$, scaling trajectories are initially dominated by the NFL fixed point, corresponding to a high temperature $T_K^{(1)}$. There is then a rapid crossover at $T_K^{(2)}$, when the effect of finite $M$ develops, to a heavy Fermi liquid (FL) fixed point with fully quenched local moments.
}
\label{scalingschem2}
\end{figure}

Now suppose we now re-introduce a small finite $M$ (or alternatively,
a departure from neutrality): this will now guarantee a Kondo effect
for any value of coupling $g$.  From a scaling perspective,
$M$ has the dimensions of energy with leading scaling behavior is 
$\frac{\partial M}{\partial \ln \Lambda} = M$, 
so that $M$
is a relevant perturbation to the Withoff-Fradkin fixed
point, and at scales lower than $M$, forcing 
conventional Kondo scaling is
re-established, scaling away to a conventional Fermi liquid strong
coupling fixed point. The schematic scaling phase diagram 
as shown in  (Fig. \ref{scalingschem2}).  

In reality, $M$ is finite and we are never precisely at particle-hole
symmetry in MATBG, nevertheless, we shall argue that for realistic
parameters, 
$g\gtappr g_{c}$, so that over a large temperature range, the physics
is dominated by the Kondo effect in a Dirac sea. 
In particular, for 
for $g>g_{c}$ and small $M$, 
the Kondo scaling trajectories 
are initially dominated by the non-Fermi liquid fixed point,
corresponding to a high Kondo temperature $T_{K}^{(1)}$. At a much
lower temperature $T_{K}^{(2)}$, the effect of finite $M$ develops,
causing the system to lose its residual entanglement entropy at a second, lower
temperature Kondo scale $T_{K}^{(2)}$.

 We can estimate $T_K^{(1)}$ by integrating the scaling equation \eqref{scalingregimes}
\begin{eqnarray}\label{TK1}
	\int_{g_{0}}^{1}\frac{dg}{g-2N_{f}g^{2}} = \int_{U/2}^{2 \pi T_{K}^{(1) }} d \ln{\Lambda}
\end{eqnarray}
where $\Lambda_{0}\sim U/2$ is the upper cutoff and $2\pi T_{K}^{(1)}$
is the lower cutoff determined by the Kondo temperature. 
This
then gives 
\begin{equation}\label{lowerTK}
T_{K}^{(1)} = \frac{U}{8 \pi (1-g_{0})}\left(1-\frac{g_{c}}{g_{0}} \right)
\end{equation}
where $g_{c}=1/2N_{f}$ and $g_{0}= \frac{4 \overline{\gamma
(k)^{2}}}{U}\rho_{c} (\Lambda_{0})$. For $M=0$, the third chiral limit (flat limit) \cite{song21, Lau23} of
MATBG, we expect that
\begin{equation}\label{TK1X}
T_{K}^{(1)}\sim \gamma,
\end{equation}
where $\gamma$ is the renormalized
hybridization gap. We will see that this is confirmed by mean-field theory.

We can also estimate $T_{K}^{(2)}$ by identifying 
with the renormalized resonant level width 
$T_{K}^{(2)}\sim \Delta$
where the renormalized hybridization $\Delta=2 \pi (\gamma/D)^{2}M$ \eqref{bwidthy}, with
$\gamma_{0}\rightarrow \gamma\sim T_{K}^{(1)}$, which gives
\begin{equation}\label{TK2}
T_{K}^{(2)}\sim \Delta\propto \pi \left(\frac{T_{K}^{(1)}}{D} \right)^{2}M \ll T_{K}^{(1)}
\end{equation}
for the Fermi liquid temperature. As in the non-interacting case,
there is a very large ratio between the two temperature scales
$T_{K}^{(1,2)}$, and in the limit $M\rightarrow 0$, non-Fermi liquid
behavior continues to zero temperature.  
The scale $T_K^{(2)}$ corresponds to the  renormalized Dirac bandwidth of the flat bands. 
The characteristic size of
the screening cloud 
\begin{equation}\label{}
\xi_{K} = \frac{v_{D}}{T_{K}^{(2)}}\rightarrow \infty 
\end{equation}
diverges in the limit $M\rightarrow 0$, reflecting the finite
entanglement entropy and the critical nature of 
the Kondo effect at the Whithoff-Fradkin fixed point. 

Normally, we would expect that a non-Fermi liquid fixed point of a
Kondo impurity would not survive in the lattice. However, the crystal
symmetries of MATBG guarantee the existence of Dirac cones at the
$K_{M}$ points which become flat when $M=0$, and as in the impurity,
this fixed point will dominate the physics over a large energy range,
guaranteeing the persistence of a two scale {\sl lattice } Kondo
effect in MATBG, allowing us to use the impurity estimates as an
order-of-magnitude guide to the corresponding scales in the lattice.
We note that the presence of a finite occupancy of the flat bands will then require the Fermi level to be pinned near the Dirac-like dispersion at the moir\' e K-points.
(We shall now show this is the case using a mean-field theory.) In
conclusion, we expect that a similar ratio between $T_{K}^{(1)}$ and
$T_{K}^{(2)}$ will persist in the Kondo lattice of MATBG, where for
$M=0$, the renormalized band-structure will contain perfectly flat
f-bands and non-Fermi liquid behavior. 


\section{Mean-field Approach}\label{sec:slaverotor}
We follow the method of Florens and Georges \cite{slaverotor1,
slaverotor2} using an auxiliary rotor description to develop a
mean-field theory. The advantage of this approach is that
it reproduces the correct strong and weak coupling limits, and is not limited
to a particular range of filling factors. enabling
us to capture the valence fluctuations and Kondo effect in MATBG at all filling factors. 

The auxiliary rotor approach follows the strategy
of earlier auxiliary boson methods
separating the physical f-electron field into a product of a spin
fermion and an ancillary charge boson, which in this case is
represented as a rotor.   The physical Hilbert space of $n=\nu+4$ f-electrons is
represented as the product of a spinon and rotor state as
follows
\begin{equation}\label{}
\vert f^{\nu+4}
\rangle = \vert \tilde{f}^{\nu+4}
\rangle \vert \nu\rangle,\qquad \qquad (\nu\in[-4,4]),
\end{equation}
where $\vert \nu\rangle $ is an angular momentum eigenstate of a rotor with
$L_{z}=\nu $, i.e. $\hat L^{z}\vert \nu\rangle=\nu\vert\nu\rangle$.  We can re-write $\vert \nu\rangle$ in the angular basis, $\langle \theta \vert \nu\rangle  = e^{i \nu \theta
}$, where 
\begin{equation}\label{}
\vert \theta \rangle  = \sum_{\nu}\vert \nu\rangle \langle \nu \vert
\theta \rangle  
= \sum_{\nu}\vert \nu\rangle e^{-i \nu \theta }.
\end{equation} 
In this representation, the physical f-electron field is separated
into a product of a spin fermion and an ancillary raising 
operator $L_{+}= \sum_{\nu}\vert \nu+1\rangle\langle
\nu\vert$, 
\begin{equation}\label{create}
f\dg _{\alpha \eta \sigma }\equiv 
\tilde{f}\dg _{\alpha \eta \sigma }L^{+},
\end{equation}
so that each time an f-fermion is added, the rotor angular
momentum increases  by one, conserving the gauge charge 
\begin{equation}\label{}
\hat  Q = \hat \nu_{f}-L^{z}, \qquad [f\dg _{\alpha \eta \sigma }, \hat Q]=0.
\end{equation}
where $\hat  \nu_f=\hat n_f-4$ is the filling factor. 
The physical Hilbert space corresponds to the gauge ``neutral'' slice of Fock-space 
where
$\hat  Q=0$, or $\nu_f=L^{z}$.
Importantly, 
the constraint is a perfect constant of motion, so that the
fluctuations in the rotor angular momentum perfectly track the
fluctuations in the physical charge. For this reason, 
merely by imposing the constraint $\langle \hat Q\rangle  =0 $, one obtains
a good approximation to the underlying Coulomb blockade physics. 

In the angular basis, 
\begin{equation}\label{}
L^{+}\vert \theta \rangle =
 \sum_{\nu}\vert \nu+1\rangle e^{-i\nu\theta } 
= \sum_{\nu}\vert \nu\rangle e^{-i (\nu-1)\theta } = e^{i\theta }\vert \theta \rangle,
\end{equation}
so that $L^{+}=e^{i\theta }$
is simply a phase factor, and we can rewrite 
physical creation operator\eqref{create} as a product of spin and charge
degrees of freedom 
\begin{equation}\label{}
f\dg _{\alpha \eta \sigma }\rightarrow 
\tilde{f}\dg _{\alpha \eta \sigma }
e^{i\theta},
\end{equation}
In this way, the phase of the rotor $\theta $ is conjugate to the
physical charge $n_{f}= - i \frac{\partial}{\partial \theta}$ of the f-state. 

With these considerations, 
the ``atomic'' interaction \eqref{phenomatom} becomes
\begin{equation}\label{}
H_{A}= \frac{\Ustar}{2}\sum_{\bR } (L^{z}_{\bR }-\kappa \nu)^{2}.
\end{equation}
while the 
hybridization \eqref{eq:thfhybridization} becomes 
\begin{eqnarray}\label{eq:thfhybridization}
	H_{fc} &=&
	\gamma_0 \sum_
{
	\textbf{R}\alpha \eta \sigma } 
\left( 
\tilde{f}^{\dagger}_{\textbf{R}\alpha \eta \sigma } 
	c_{\textbf{R} \alpha \eta \sigma  }L_{{\bR }}^{+}  + \text{h.c.}\right).
\end{eqnarray}
The mean-field theory is obtained by imposing $\langle  Q\rangle =0$
with a Lagrange multiplier, treating the rotor and fermionic
degrees of freedom as separate degrees of freedom subject to this constraint.
With this approximation, the hybridization is renormalized
\begin{eqnarray}\label{constraint1}
\gamma_{0}\rightarrow {\gamma} = 
\gamma_{0}\langle L^{+}_{\bR }\rangle 
=\gamma_{0}\langle e^{i\theta }\rangle \rightarrow 
\gamma_{0}\langle \cos \theta \rangle,\\ \phantom{.} \nonumber
\end{eqnarray}
where for convenience, one chooses a phase where $\langle \sin \theta
\rangle =0$. The coupling between the fermions and the rotor defined
by $H_{fc}$ produces a transverse Weiss field on the rotor, giving
rise to a mean-field rotor Hamiltonian of the following form
\begin{equation}\label{}
H_{rot}= \frac{\Ustar}{2} (L_{z}-\kappa \nu)^{2}- K \cos\theta 
\end{equation}
subject to the constraint 
\begin{equation}\label{constraint2}
K = - 2\gamma_{0}\langle \tilde{f}\dg_{\bR
\alpha \eta \sigma }c_{\bR \alpha \eta \sigma }\rangle  .
\end{equation}

\begin{widetext}
We now reformulate the mean-field theory as a variational Hamiltonian, 
rewriting the lattice Hamiltonian \eqref{Andersonlattice} as,
\begin{eqnarray}
	H &=& H_{\text{rot}} + H_F + N_s \left(\frac{\gamma K}{\gamma_0} + \mu \nu\right) \\
	H_{\text{rot}} &=& N_s \left[\frac{\Ustar}{2} L_{z}^2 - (\xi + \Ustar \kappa \nu) L_{z} - K \cos{\theta}\right] \\
	H_F &=& \sum_{\bk \eta \sigma } \left[\Psi\dg_{\bk \eta \sigma}
	\textbf{$\mathcal{H}$}_{\text{F}}\left(\bk\right) \Psi_{\bk
	\eta \sigma}\right] + \mu \langle N_c\rangle_{\nu = 0} - N_s(\xi - \mu) N_f,
\end{eqnarray}
where $\gamma$, $K$ and $\xi$ are determined by the stationary
points of the mean-field free energy. 
Note that the cross-term $N_{s}\frac{\gamma K}{\gamma_{0}}$ in
$H$ results from a Hubbard-Stratonovich decoupling of the
hybridization \eqref{eq:thfhybridization}, permiting
us
to vary $\gamma$ and $K$ independently, using stationarity to impose the
constraints \eqref{constraint1} and \eqref{constraint2}.
The matrix
\begin{equation}\label{fermionichamiltonian}
	{\cal H}_{F} (\bk )= \pmat{ (\xi -\mu)\sigma_{0}
	&  \gamma \phi^{(\eta) }
(\bk) 
& \dots & \gamma \phi^{(\eta) }(\bk +\bG_n)\cr
\gamma \phi^{(\eta) }(\bk)\dg
 &  {\cal
	H}^{(\eta)}(\bk)
	- \mu{\underline{1}} &0 &0 \\
	\vdots &0 & \ddots & 0 \\ 
\gamma \phi^{(\eta) }
(\bk+\bG_{n})\dg  & 0 & \dots & {\cal H}^{(\eta)}(\bk + \bG_n)
	- \mu{\underline{1}}}
\end{equation}
determines the quasiparticle dispersion with renormalized hybridization strength $\gamma$, where the
	conduction electron dispersion matrix ${\cal H}^{\eta } (\textbf{k})$
are  defined in \eqref{eqx:conductiondispersion}
and the hybridization matrices 
$\phi^{(\eta )} (\bk )$ defined in
\eqref{eq:hybridizationmatrix}. $\langle N_c \rangle_{\nu = 0}$ is
the number of c-electrons at half-filling and $N_f = 4$ is the valley
spin degeneracy. $L_{z}$ is the angular momentum operator. 
We rewrite the rotor Hamiltonian as 
	\begin{equation}\label{rotorhamiltonian}
	H_{\text{rot}} = \frac{\Ustar}{2}L^2_z - (\xi + \Ustar \kappa \nu) L_z - \frac{K}{2}\left(L_+ + L_-\right).
\end{equation}
	$\xi$ is the Lagrange multiplier that constrains the allowed values of the angular momentum component $L_z = m_l \in [-N_f, N_f]$.
\begin{equation}\label{}
	\Psi _{\bk \eta \sigma} = (\tilde{f}_{\bk 1\eta
	\sigma},\tilde{f}_{\bk 2 \eta \sigma}, c_{\bk 1\eta  \sigma},
c_{\bk 2\eta  \sigma}, 
c_{\bk 3 \eta  \sigma}, c_{\bk 4\eta  \sigma},c_{\bk+\bG_1 1\eta
\sigma}, \dots, c_{\bk+\bG_n 3\eta  \sigma}, c_{\bk+\bG_n 4 \eta
\sigma})^{T}
\end{equation}
is a spinor combining the four conduction fields at each reciprocal lattice vector
and the two f-electron operators at each valley $\eta =\pm 1$ and spin
$\sigma=\pm 1$. (For convenience, henceforth we will drop the tilde on the f-fields.) Notice that while $H_{MF} (\bk )$ commutes with the spin
and valley quantum numbers, at general momentum it breaks the two-fold
$\Gamma_{3}$ degeneracy down to a $N_f=4$ fold valley-spin degeneracy. 

The mean-field Free energy per unit cell obtained by integrating out the fermions
for a static configuration of the fields $(\gamma,\xi, K)$, 
is then 
\begin{eqnarray}\label{eq:mfhalffillfreeenergy}
	\mathcal{F} 
	&=& F + \Phi_{\text{rot}} + \left(\frac{\gamma K}{\gamma_0} + \mu \nu\right)\\
	F &=& - \frac{N_f}{N_s} T \sum_{\bk }{\rm Tr}
 {\rm \ ln} (1 + e^{- \beta {\cal H}_{F} (\bk )})
	+ \mu  \langle N_c\rangle_{\nu = 0} - (\xi - \mu) N_f \\
	\Phi_{\text{rot}} &=& - T {\rm \ ln} {\rm Tr} \left[e^{-\beta H_{\text{rot}}}\right]
\end{eqnarray}
Writing  
\begin{eqnarray}\label{statmech}
-T  N_{f}\sum_{\bk } {\rm Tr}[1 + e^{-\beta  {\cal H}_{F} (\bk )}] &=& -T N_{f}\sum_{\bk } {\rm Tr}[2 \cosh
(\beta  {\cal H}_{F} (\bk )/2)] + \frac{ N_{f}}{2}\sum_{\bk }{\rm Tr}{\cal  H}_{F} (\bk )
\cr
&=& -T  N_{f}\sum_{\bk }{\rm Tr}[2 \cosh
	(\beta  {\cal H}_{F} (\bk )/2)] + N_{s}N_{f}\bigl[(\xi - \mu) -\mu 
 2 N_{G}\bigr]\cr
&=& -T N_{f}\sum_{\bk }{\rm Tr}[2 \cosh
	(\beta  {\cal H}_{F} (\bk )/2)] + N_{s}(\xi-\mu)N_{f} - \mu 
 \langle N_c\rangle_{\nu = 0}.
\end{eqnarray}
where $N_{G}$ ($N_{G}$=7 in our calculations) is the number of reciprocal lattice vectors  $\vec{G}$, including
the origin included in ${\cal H}_{F}$, we can rewrite the mean-field
free energy per unit cell as
\begin{eqnarray}\label{eq:mfparticlehole}
	\mathcal{F} 
	= - \frac{N_f}{N_s} T \sum_{\bk }{\rm Tr}
 {\rm \ ln} (2 \cosh  (\beta {\cal H}_{F} (\bk )/2))
	+ \Phi_{\text{rot}} + \left(\frac{\gamma K}{\gamma_0} + \mu \nu\right).
\end{eqnarray}
\end{widetext}

 The saddle-point requirement that $F$ be
stationary with respect to variations in $K$, $\gamma$, and $\xi$ imposes the coupled self-consistency conditions
\begin{equation}\label{eq:K-selfconsistent}
	\langle \cos \theta \rangle_{\theta} = \frac{\gamma}{\gamma_0},
\end{equation}
\begin{equation}\label{eq:gamma-selfconsistent}
	K = -2\frac{\gamma_0}{N_s }\sum_
	{\bR, B 
	}
	\langle c^{\dagger}_{\bR B}\tilde{f}_{\textbf{R}B } 
	\rangle,
\end{equation}
and
	\begin{eqnarray}\label{eq:lambda-sc}
		\langle L_z\rangle &=& \frac{1}{N_s} \sum_{\bR, B} \langle \tilde{f}^{\dagger}_{\textbf{R}B} \tilde{f}_{\textbf{R}B} \rangle - N_f \\ \nonumber &=& \nu_f .
\end{eqnarray}
Stationarity of the free energy with respect to the chemical potential $\mu$ fixes the total number of electrons $N_e$ in the system,
\begin{eqnarray}\label{eq:mu-sc}
	-\frac{\partial \mathcal{F}}{\partial \mu} = \nu_c + \nu_f -\nu
	=0,
\end{eqnarray}
where $\nu =  (N_e - \langle N_e\rangle_{\nu = 0})/N_s $ is the filling factor and we have denoted $\nu_c=
(\langle \hat  N_{c}\rangle - \langle \hat N_{c}\rangle_{\nu=0})/N_{s}
$ .


The key scales of the mean-field mixed valent moir\'e lattice model for TBG are set by the impurity physics before coherence is reached. Hence, the doping-temperature phase diagram for the mixed valent moir\'e lattice model would greatly resemble a Doniach phase diagram based on a single impurity model for the f-states in MATBG. 

The mean-field hybridization width for the mixed valent moir\'e lattice is
\begin{equation}\label{eq:meanfieldwidth}
	\Delta[\mu] = \pi \rho_c(\mu)  \overline{\gamma^{2}(k)},
\end{equation}
where $\overline{\gamma^{2}(k)} = (\gamma^2/\Gstar)\, \overline{\gamma_0^{2}(k)}$ is the momentum integrated average of the mean-field hybridization squared. We anticipate the approximate mean-field bandwidth of the flat band to
be 
\begin{equation}\label{MFbandwidth}
	\tilde{W}_{\text{MF}} =  v_D^{\text{MF}} K_{\theta}  = \frac{\gamma^2}{\Gstar} W \sim \Delta,
\end{equation}
where $v_D^{\text{MF}}$ is the Dirac velocity at the $K_M$ points for the
mean-field theory, and $W$ is the bare noninteracting bandwidth \eqref{barescales}. 


\section{Mean-field Results}\label{sec:meanfieldresults}
\begin{figure}[h]
	\centering
	\includegraphics[width = \columnwidth]{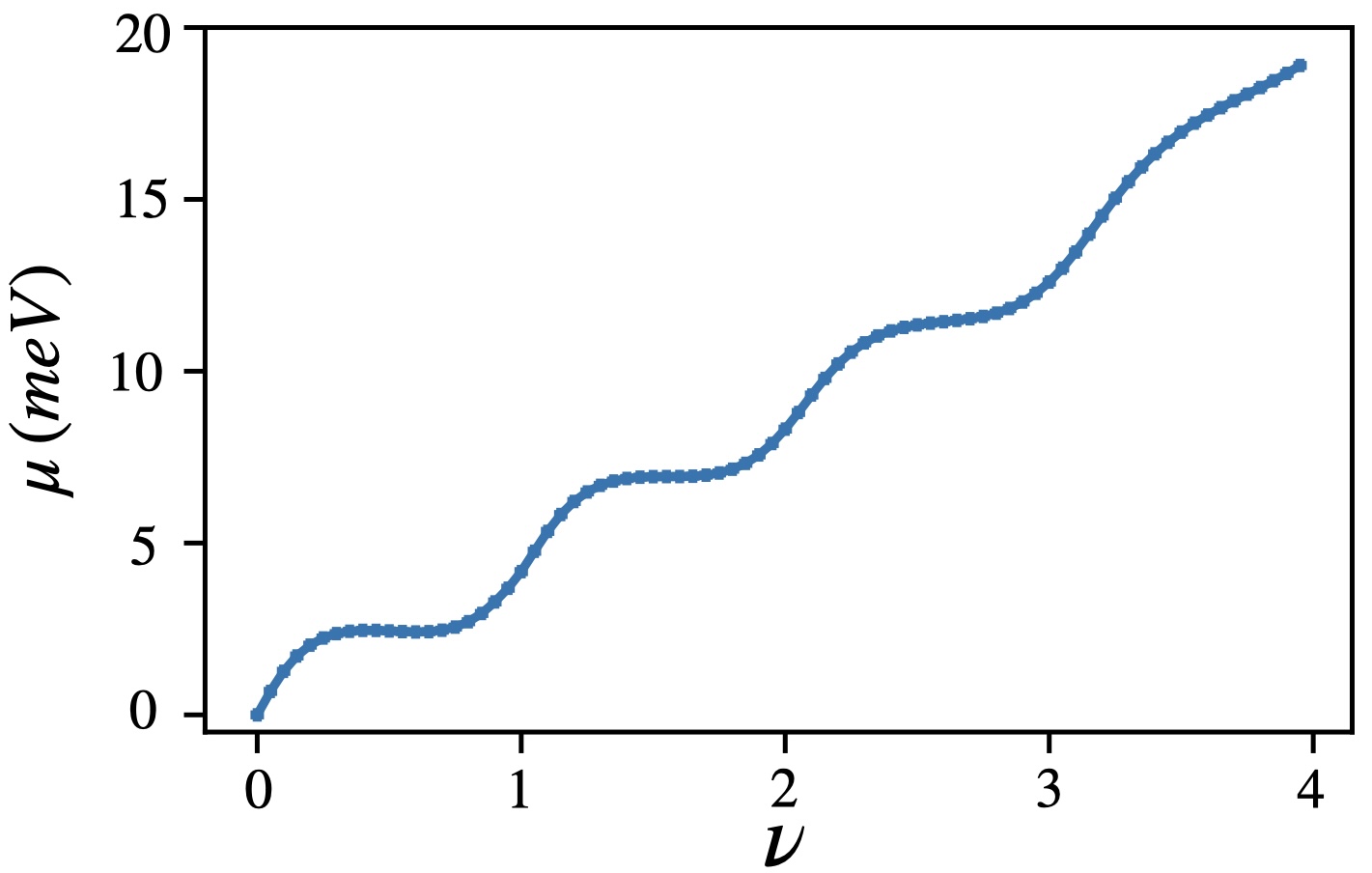}
	\caption{Chemical potential $\mu$ variation with total filling factor $\nu$ for $T = 0.1 \, meV$ with phenomenological parameters $\Ustar = 30 \, meV$ and $\kappa = 0.8$ within the mean-field theory. Like the ``atomic'' limit, the chemical potential jumps at the integer filling factors and partially resets between successive integer fillings. The analytic chemical potential of the jump by $\Ustar$ and the partial reset by $\Ustar \kappa$ within the ``atomic'' limit are renormalized by both finite hybridization and finite temperature effects. }
	\label{mureset}
\end{figure}
In our calculations, we have adopted the phenomenological values
$\Ustar = 30 \, meV$,  $\kappa = 0.8$ and $\gamma_{0}=  6.5$meV. 
The choice of $\gamma_0$ 
is a delicate balance: if $\gamma_0$ is too large then no local
moment behavior survives and the chemical potential has no reset
behavior; if $\gamma_0$ is too small, then the bandgap between the
flat bands and remote bands at $\Gamma_M$ never fully opens and the
flat bands never become topological. Our choice of $\gamma_0$
preserves the resetting behavior of the chemical potential $\mu[\nu]$
as a function of filling, at the cost of the gap to the remote bands
from opening.\cite{foot1}

\begin{figure}[t]
	\centering
	\includegraphics[width = \columnwidth]{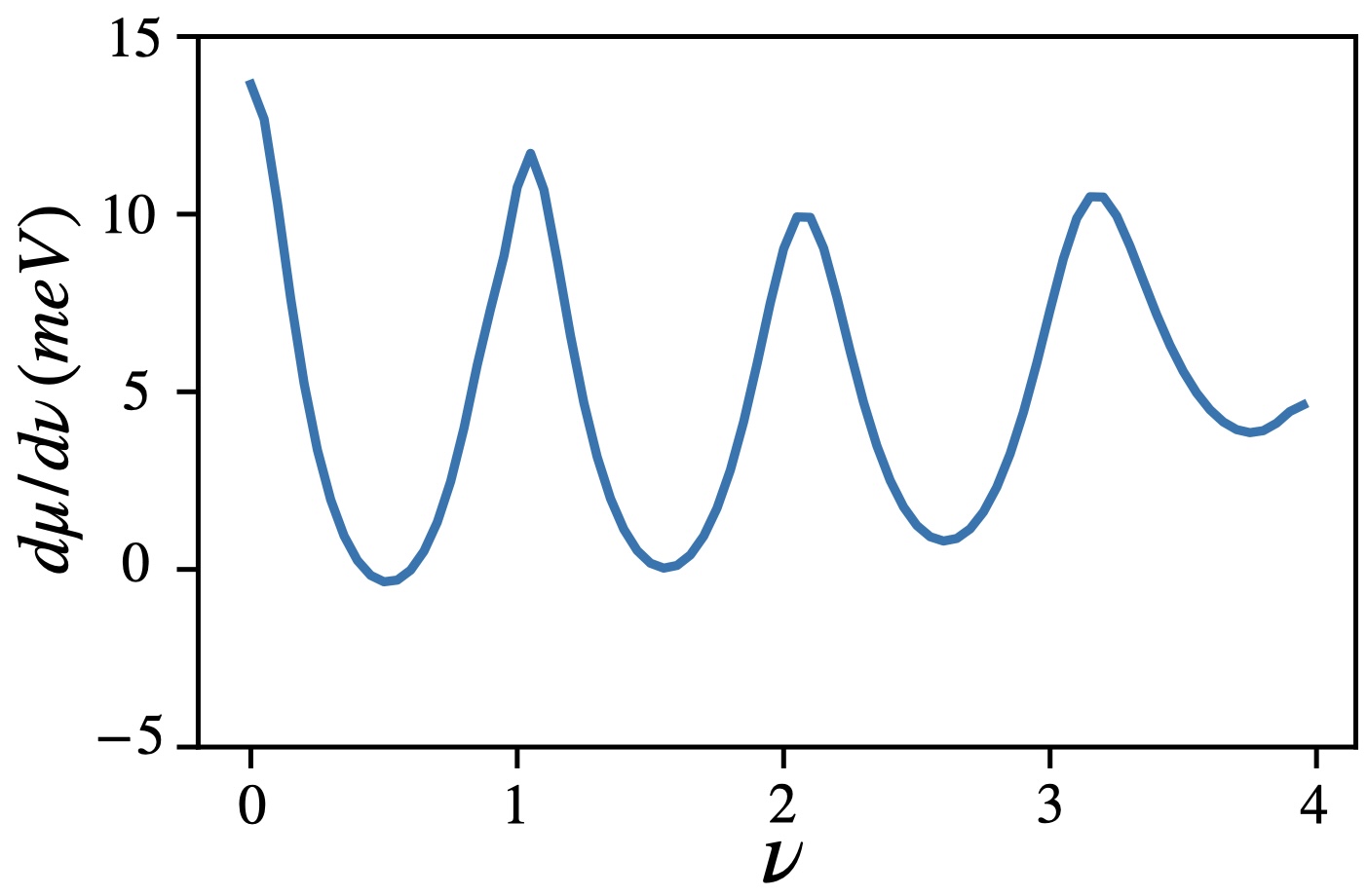}
	\caption{Inverse compressibility $d\mu/d \nu$ variation with total filling factor $\nu$ for $T = 0.1 \, meV$ with phenomenological parameters $\Ustar = 30 \, meV$ and $\kappa = 0.8$ within the mean-field theory. Like the ``atomic'' limit, we get small negative values around half integer fillings. The scales from the ``atomic'' limit are renormalized by both finite hybridization and finite temperature effects. }
	\label{inversecompressibility}
\end{figure}

With these reduced bare parameters, the 
non-interacting bandwidth \eqref{barebandwidth} is reduced to,
\begin{equation}\label{barescales}
	W \approx { 0.02 M \sim 1.0 {\rm K}}
\end{equation}
We then numerically solve the self consistent equations
\ref{eq:gamma-selfconsistent}, \ref{eq:lambda-sc}, and \ref{eq:mu-sc}
as a function of filling factor. The particle-hole symmetric results are
displayed for positive doping. 
With the choice of parameters above, our mean-field theory reproduces
the ``atomic'' limit (sec. \ref{coulombblockade}). The chemical potential
$\mu$ undergoes a partial reset as a periodic function of filling
(Fig. \ref{mureset}) and the inverse compressibility has peaks at integer fillings (Fig. \ref{inversecompressibility}), a consequence of the Coulomb blockade physics
and the emergent f-electron potential. The scales from the analytic unhybridized ``atomic'' limit (Fig. \ref{phenom} a) are renormalized in our mean-field results by both the finite hybridization and finite temperature effects.

\subsection{Unquenched f-states}
\begin{figure}[b]
	\centering
	\includegraphics[width =\columnwidth]{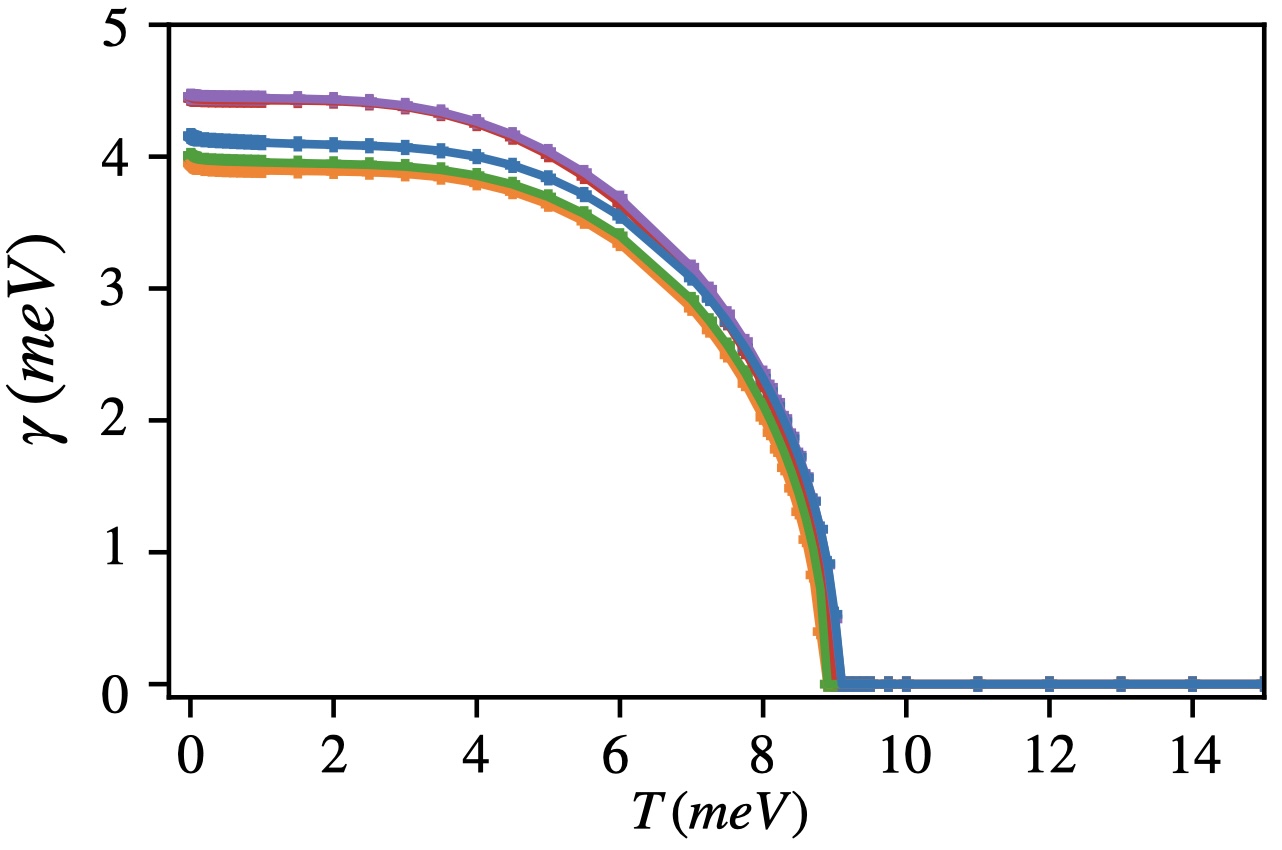}
	\caption{Mean-field order parameter $\gamma(T)$ plotted against temperature for fillings $\nu = 0$ (orange), $0.5$ (red), $1$ (green), $1.5$ (purple), and $2$ (blue). We find the Kondo order transition $T_K^{(1)} \sim 9 \, meV$. Bare hybridization used: $\gamma_0 = 6.5 \, meV$.}
	\label{gammawarmup}
\end{figure}

We observe development of nonzero hybridization $\langle \cos{\theta}\rangle_{\theta} = \gamma/ \gamma_0 \neq 0$ at a characteristic temperature $T_K^{(1)} \approx 9 \, meV$ (Fig. \ref{gammawarmup}), formally indicating the onset of the Kondo effect. This would typically signal the fractionalization of the local moments into heavy f-states and the formation of a heavy Fermi Liquid state below $T_K^{(1)}$ in conventional heavy fermion systems.

However, the relevant energy scale for Fermi liquid formation
is the much smaller f-band width $\tilde{W}_{\text{MF}}$ \eqref{MFbandwidth}, which gives rise to a much lower coherence temperature $T_K^{(2)} \sim \tilde{W}_{\text{MF}} \ll T_K^{(1)}$. This disparity leads to the persistence of thermally active, unquenched f-states in the wide intermediate temperature range between $T_K^{(1)}$ and $T_K^{(2)}$. 
The signatures of these unquenched f-states include a
Curie-Weiss magnetic susceptibility down to $T_K^{(2)}$, 
local moments with an unquenched entropy that persists below 
$T_K^{(1)}$ and ultimately quenching below $T_K^{(2)}$, giving rise to a specific heat feature around $T_K^{(2)}$.

We calculate the magnetic-susceptibility $\chi = -\partial^2
\mathcal{F}/ \partial B^2$, evaluated at $B = 0$ from the bubble
diagram of the f-spinors, which is 
is equivalent to calculating $\left .-\partial^2 \mathcal{F}/ \partial
\mu^2\right\vert_\xi$ at a fixed constraint field $\xi$. 
At high temperatures larger than the Kondo temperature, we expect the
magnetic-susceptibility to assume a Curie-Weiss form, 
\begin{equation}\label{curieweissform}
	\chi(T) \propto \frac{m^2}{T + \theta},
\end{equation}
where $m$ is the magnetic moment of each free local moment, and $\theta$ is the Curie-Weiss temperature, a phenomenological scale taking care of interactions between moments.


\begin{figure}[t] \centering \includegraphics[width
=\columnwidth]{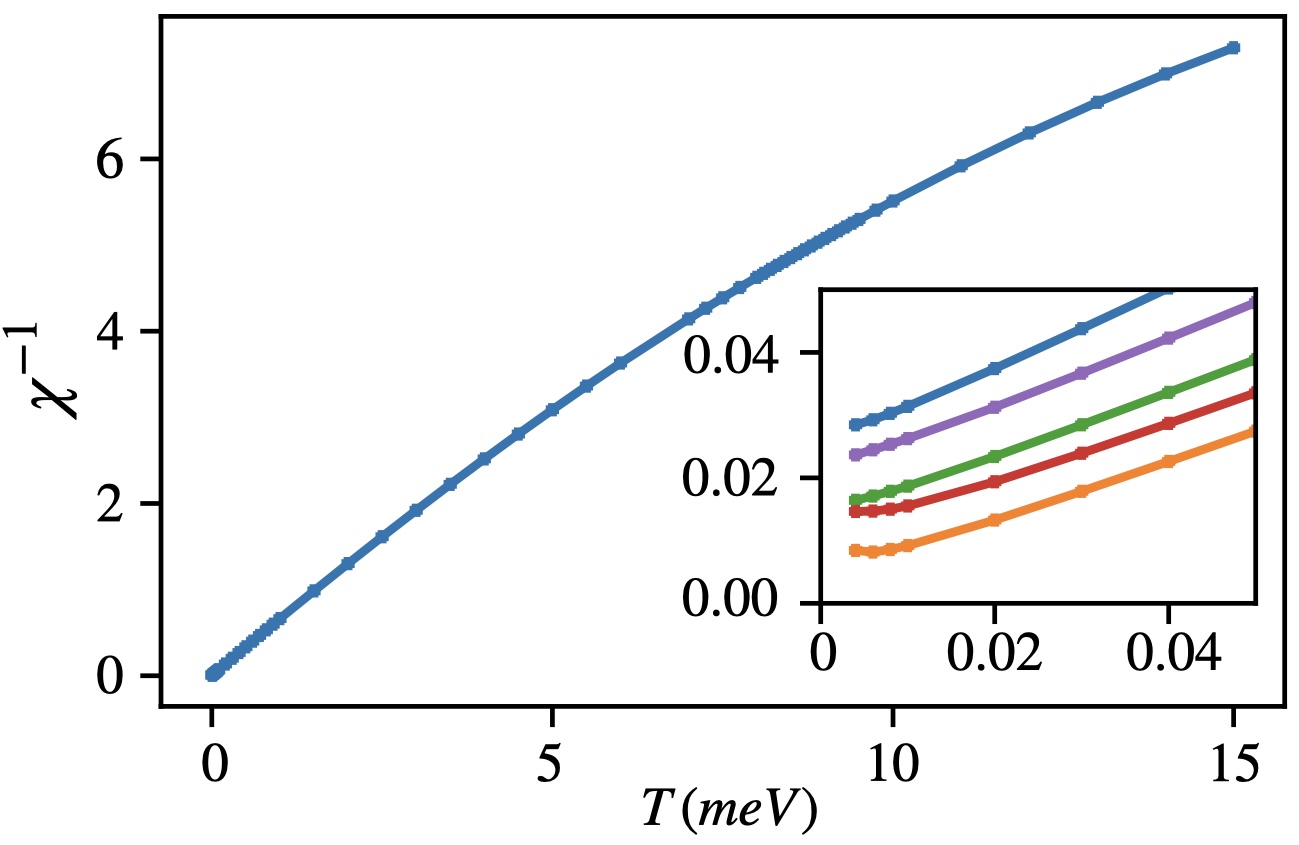} \caption{Inverse spin
and valley susceptibility $\chi^{-1}$ as a function of temperature $T$
for integer filling $\nu = +2$ plotted in blue. There is a subtle crossover
between two Curie-Weiss behaviors around the temperature
$T_K^{(1)} \approx 9 \, meV$ where Kondo order turns on, (Inset)
Inverse spin/valley susceptibility for fillings $\nu = 0$ (orange), $0.5$
(red), $1$ (green), $1.5$ (purple), and $2$ (blue), offset by
different constants for clarity. Curie-Weiss behavior persists well
below the Kondo ordering temperature $T_K^{(1)}$ and magnetic $SU(8)$
local moments in MATBG are only fully screened at around $T_K^{(2)}
\sim 0.01 meV$ where Curie-Weiss behavior ends. } \label{curieweiss}
\end{figure}

Typically in heavy fermion systems, the onset of Kondo order (Fig. \ref{gammawarmup}) at $T_K^{(1)}$ coincides with the end of Curie-Weiss magnetic-susceptibility due to Kondo screening. Intriguingly, and distinct from conventional heavy fermion phenomenology, our mean-field theory for MATBG finds that the Curie-Weiss behavior, characterized by the linear temperature dependence of the inverse magnetic-susceptibility $\chi^{-1}$, persists below the Kondo ordering temperature $T_K^{(1)}$ down to a much smaller Kondo coherence scale $T_K^{(2)}$, (Fig. \ref{curieweiss} inset). 


At the mean-field condensation temperature $T_K^{(1)}$, we observe a transition between two different Curie-Weiss behavior in the inverse magnetic susceptibility $\chi^{-1}(T)$, exhibiting an increased gradient below the mean-field ordering temperature $T_K^{(1)}$ (Fig. \ref{curieweiss}). The gradient of the inverse magnetic-susceptibility is inversely proportional to the square of the magnetic moment of the free local moments \eqref{curieweissform}. Hence, the gradient increase indicates a partial quenching of the $SU(8)$ local moments at $T_K^{(1)}$.

To verify the partial screening of the $SU(8)$ local moments at the mean-field transition temperature $T_K^{(1)}$ and their thermal nature until the f-states are fully quenched at the Kondo coherence temperature $T_K^{(2)}$, we calculate the entropy of the f-state,
\begin{eqnarray}\label{entropyanalytic}
	S = &-& \sum_{n} \left(|\langle n | 1_f \rangle|^2 + |\langle n | 2_f \rangle|^2 \right) \times \cr &&\left(f_{n} \ln{ f_{n}} + (1- f_{n}) \ln{(1-f_{n})}\right)
\end{eqnarray}
by projecting the analytic entropy for the eigenstate $| n \rangle$ of the fermionic mean-field Hamiltonian $\mathcal{H}_F$ \eqref{fermionichamiltonian}, onto $|1_f\rangle$, and $|2_f\rangle$, the two orbital eigenstates of the f-state in the moir\'e ``atomic'' limit with zero hybridization to the c-electrons. In \eqref{entropyanalytic}, $f_{n} = f(E_{n}, T)$ is the fermi-dirac distribution.  
\begin{figure}[b]
	\centering
	\includegraphics[width =\columnwidth]{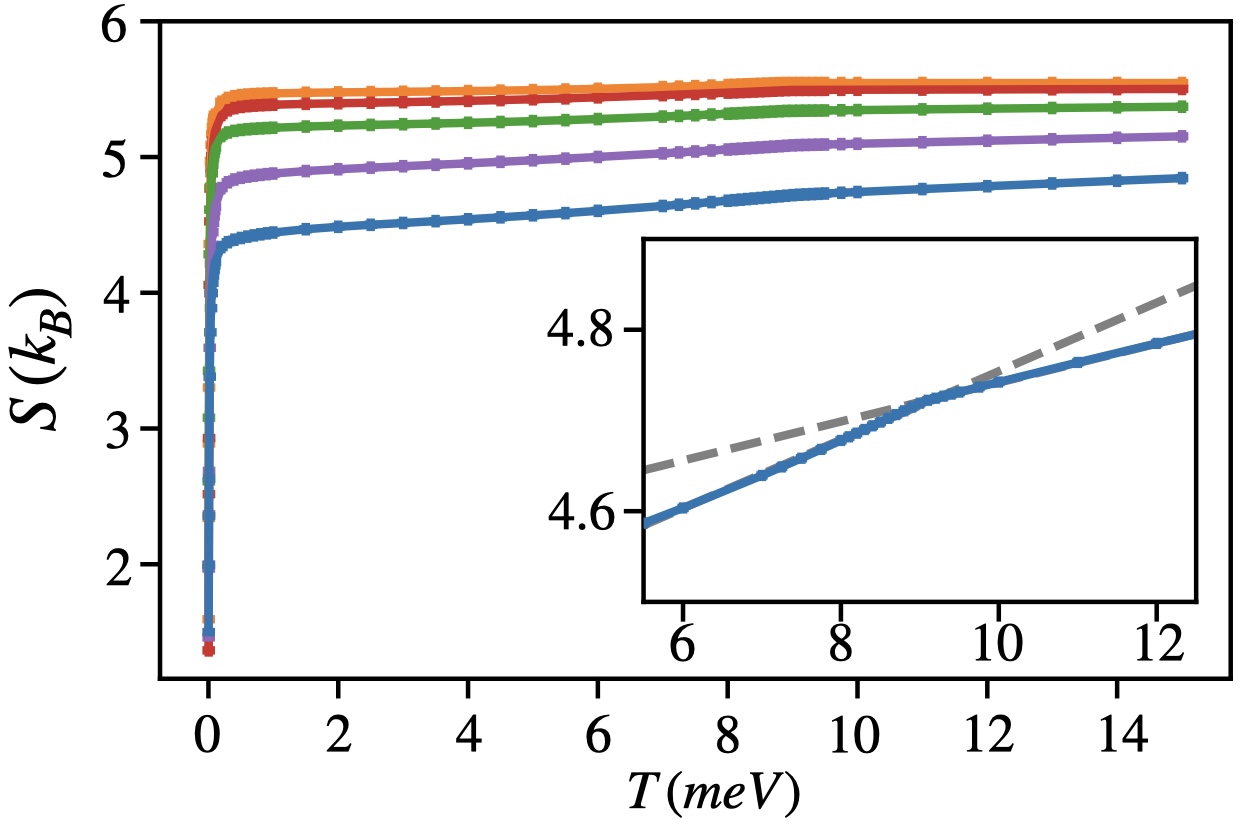}
	\caption{Entropy $S$ of the f-states as a function of temperature $T$ for fillings $\nu = 0$ (orange), $0.5$ (red), $1$ (green), $1.5$ (purple), and $2$ (blue). The entropy between $T_K^{(1)} \approx 9 \, meV$ and $T_K^{(2)} \sim 0.01 \, meV$ has good agreement with the high temperature entropy $S_{\text{therm}} \approx 4.5 k_B$ for a thermal state with 6 out of 8 f-states filled. Inset of a zoomed in view around the mean-field transition temperature $T_K^{(1)}$, dotted lines show a gradient decrease in the f-state entropy just below $T_K^{(1)}$. }
	\label{entropy}
\end{figure}

\begin{figure}[b]
	\centering
	\includegraphics[width =\columnwidth]{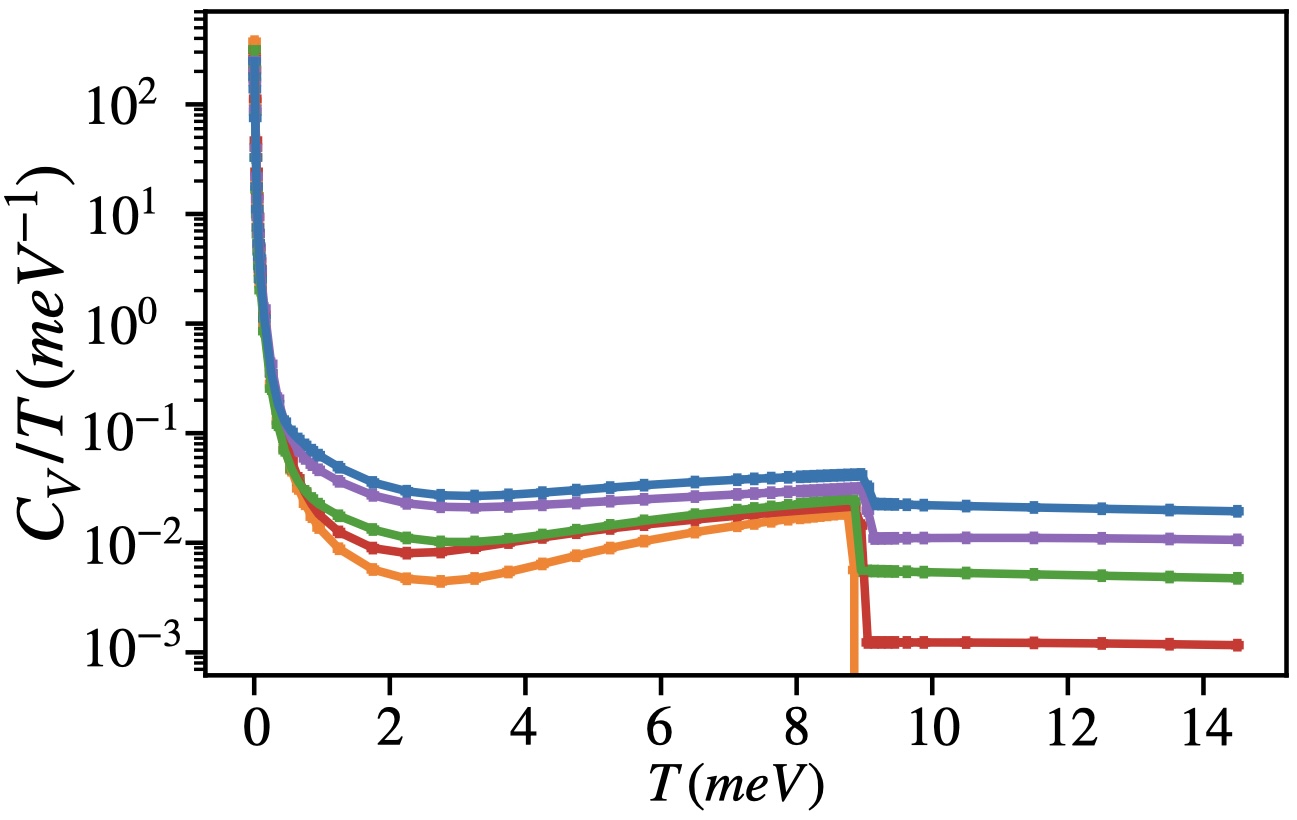}
	\caption{Specific heat capacity divided by temperature $C_V/T$ as a function of temperature $T$ for fillings $\nu = 0$ (orange), $0.5$ (red), $1$ (green), $1.5$ (purple), and $2$ (blue). The discontinuous jump at $T_K^{(1)} \approx 9 \, meV$ is a signature of the mean-field second order phase transition for the Kondo phase. At the Kondo coherence temperature $T_K^{(2)} \sim 0.01 \, meV$, $C_V/T$ peaks at a large finite value, a signature of the large effective mass of the quasiparticle carriers when coherence has formed.}
	\label{fig:specificheat}
\end{figure}
 We confirm the incoherent thermal nature of the f-states in the intermediate temperature regime $T_K^{(2)} < T < T_K^{(1)}$ by comparing the f-state entropy with the high temperature fermionic entropy of a thermal state for $\nu = 2$, where $p = 6/8$ of the 8 f-states are filled,
\begin{eqnarray}\label{thermalentropy}
	S_{\text{therm}} &=& - 8 k_B \left(p \ln{p} + (1 - p) \ln{(1-p)} \right) \cr
	&\approx& 4.5 k_B.
\end{eqnarray}

Our numerical results (Fig. \ref{entropy}) also show a good agreement with the thermal fermionic entropy \eqref{thermalentropy} between $T_K^{(1)}$ and $T_K^{(2)}$ for the other filling factors, providing strong evidence that the f-states remain thermalized and incoherent below the mean-field transition $T_K^{(1)}$, all the way down to the Kondo coherence temperature $T_K^{(2)}$.

We further calculate the f-state specific heat to temperature ratio by taking the temperature derivative of the f-state entropy $S$,
\begin{equation}\label{specificheat}
	\frac{C_V}{T} = \frac{d S}{d T}.
\end{equation}
At low temperatures, $C_V/T$ is proportional to the effective mass of the quasiparticle carriers, in our numerical results, $C_V/T$ peaks at a finite large value around the Kondo coherence temperature $T_K^{(2)}$ (Fig. \ref{fig:specificheat}), indicating that the coherent quasiparticle carriers are heavy.

\subsection{Valence Fluctuations}
To show that our mean-field theory captures the valence fluctuations in the model, we calculate the distribution of valences as a function of filling. We do this by calculating the overlap between the ground state $| 0_{\text{rot}} \rangle$ of the rotor Hamiltonian \eqref{rotorhamiltonian} using the mean-field parameters $\gamma$, $\lambda$, and $K$ solved for self-consistently at $T = 0.1 \, meV$, 

\begin{equation}
	P(Q) = |\langle Q | 0_{\text{rot}} \rangle |^2,
\end{equation}
where $| Q \rangle $ is the angular momentum eigenstate corresponding to valence $Q$ and the probabilities are  normalized such that $\sum_Q P(Q) = 1$. In the Kondo limit, $P(Q = 4 + \nu) = 1$, with no mixed valency.

\begin{figure}[t]
	\centering
	\includegraphics[width =\columnwidth]{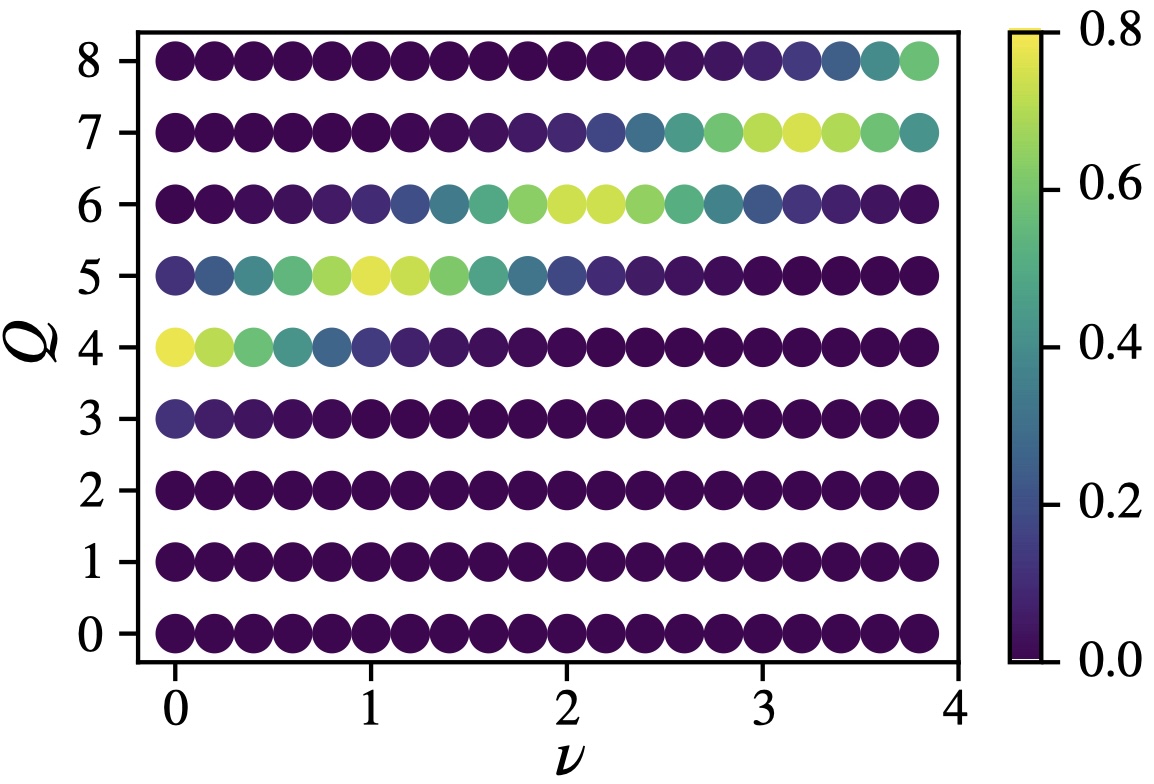}
	\caption{Valence $Q$ distribution of the f-electrons as a function of filling $\nu$, the colorbar represents the probability of finding valence state $Q$ at filling $\nu$. Parameters used: $\Ustar = 30 \, meV$, $\kappa = 0.8$ with $\gamma_0 = 6.5\,meV$ at $T = 0.1 \, meV$.}
	\label{valencedist}
\end{figure}

As shown in Fig. \ref{valencedist}, we observe a sharply peaked valence distribution at integer filling factors. In contrast, away from the integer filling factors, the probability of populating neighboring valence states (e.g. $Q = 3,5$ for $\nu = 0$) becomes comparable to the probability of the central valence $\langle n_{f \bR} \rangle = 4 + \nu$. This  finding indicates the presence of strong valence fluctuations around half-integer filling factors, despite the established of Kondo order at temperatures below $T_K^{(1)}$.

This mixed valence behavior departs from the prototypical Kondo limit, where a single well-defined valence state would dominate. 
\begin{figure}[b]
	\centering
	\includegraphics[width =\columnwidth]{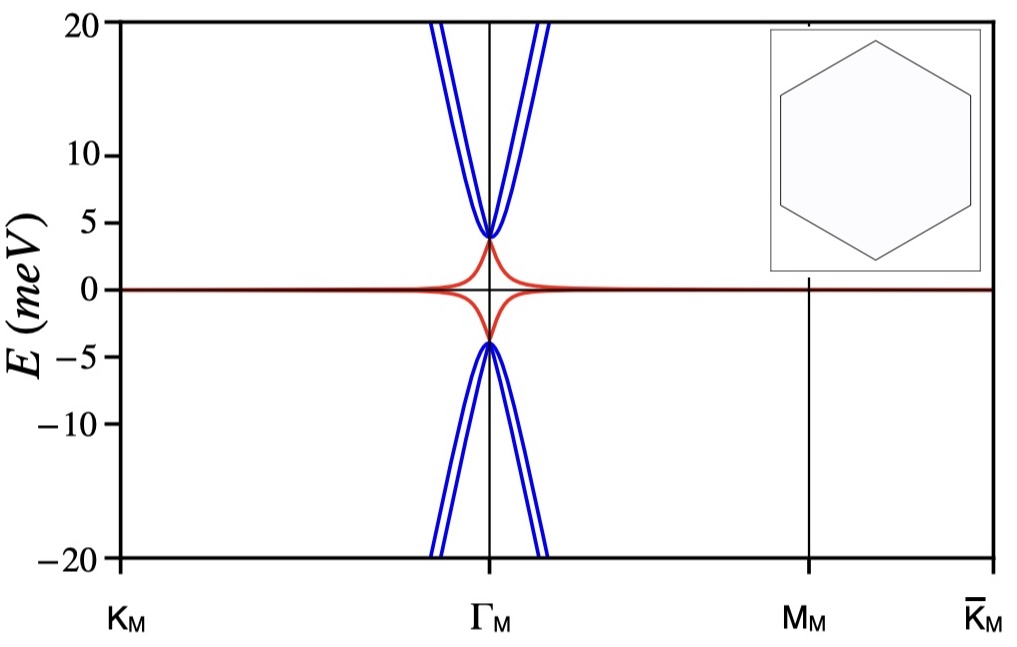}
	\caption{Many-body mean-field band-structure for a half filled flat band $\nu = 0$. Parameters used: chemical potential  $\mu = 0\, meV$, $\xi =0\, meV$, $\Ustar = 30\, meV$, $\kappa = 0.8$, with $\gamma_0 = 6.5 \, meV$, and $\gamma = 3.96\, meV$ at $T = 0.004 meV$. Inset shows Brillouin zone with no electron or hole pockets for half filling.}
	\label{halffillbs}
\end{figure}
\subsection{Quasiparticle Banstructures}
\begin{figure}[t]
	\centering
	\includegraphics[width = \columnwidth]{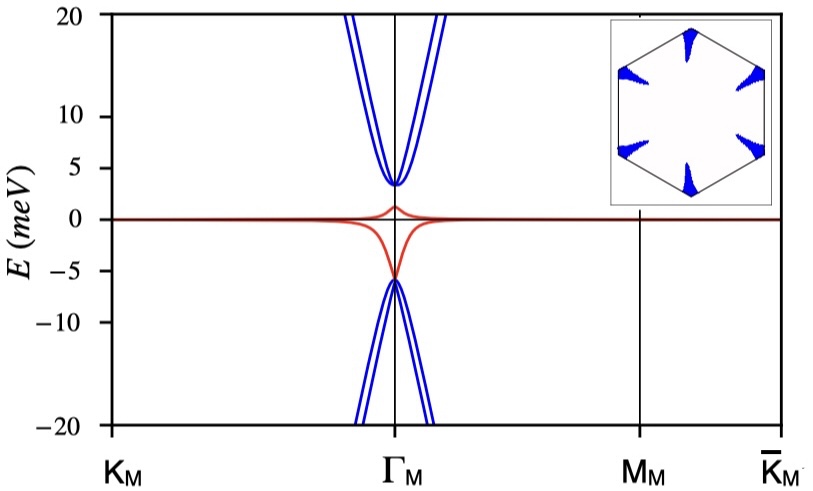}
	\caption{Many-body mean-field band-structure for $\nu = +0.5$. Parameters used: chemical potential  $\mu = 2.46\, meV$, $\xi = 2.47\, meV$, $\Ustar = 30\, meV$, $\kappa = 0.8$, with $\gamma_0 = 6.5 \, meV$, and $\gamma = 4.45\, meV$ at $T = 0.004\, meV$. Inset shows Brillouin zone with electron pocket relative to half-filling, shaded in blue.}
	\label{nu05bs}
\end{figure}

\begin{figure}[t]
	\centering
	\includegraphics[width = \columnwidth]{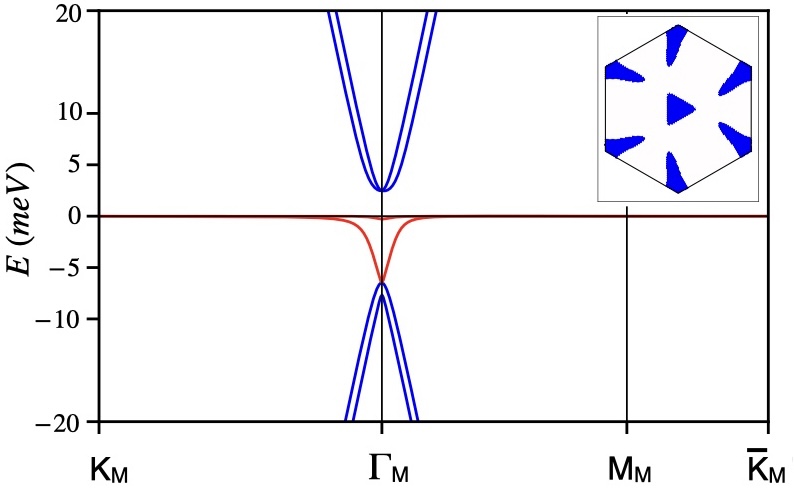}
	\caption{Many-body mean-field band-structure for $\nu = +1$. Parameters used: chemical potential  $\mu = 3.99\, meV$, $\xi = 4.01\, meV$, $\Ustar = 30\, meV$, $\kappa = 0.8$, with $\gamma_0 = 6.5 \, meV$, and $\gamma = 4.01\, meV$ at $T = 0.004\, meV$. Inset shows Brillouin zone with electron pocket relative to half-filling, shaded in blue.}
	\label{nu1bs}
\end{figure}

\begin{figure}[t]
	\centering
	\includegraphics[width = \columnwidth]{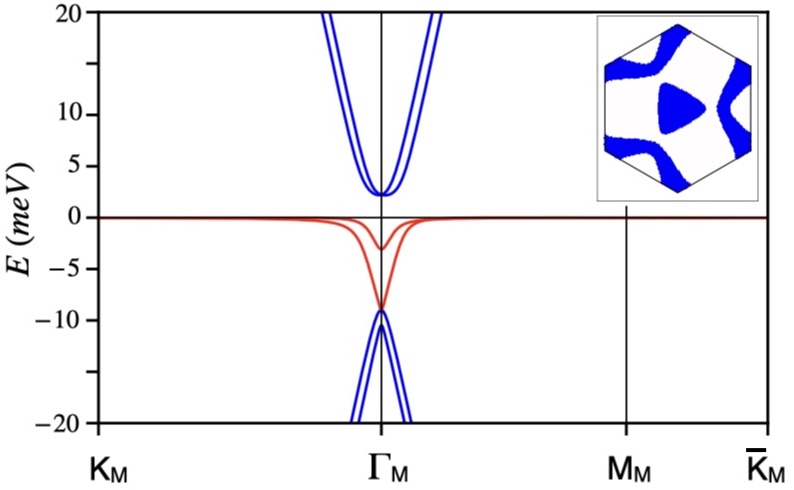}
	\caption{Many-body mean-field band-structure for $\nu = +1.5$. Parameters used: chemical potential  $\mu = 6.77\, meV$, $\xi = 6.82\, meV$, $\Ustar = 30\, meV$, $\kappa = 0.8$, with $\gamma_0 = 6.5 \, meV$ and $\gamma = 4.46\, meV$ at $T = 0.004\, meV$. Inset shows Brillouin zone with electron pocket relative to half-filling, shaded in blue.}
	\label{nu15bs}
\end{figure}

\begin{figure}[t]
	\centering
	\includegraphics[width = \columnwidth]{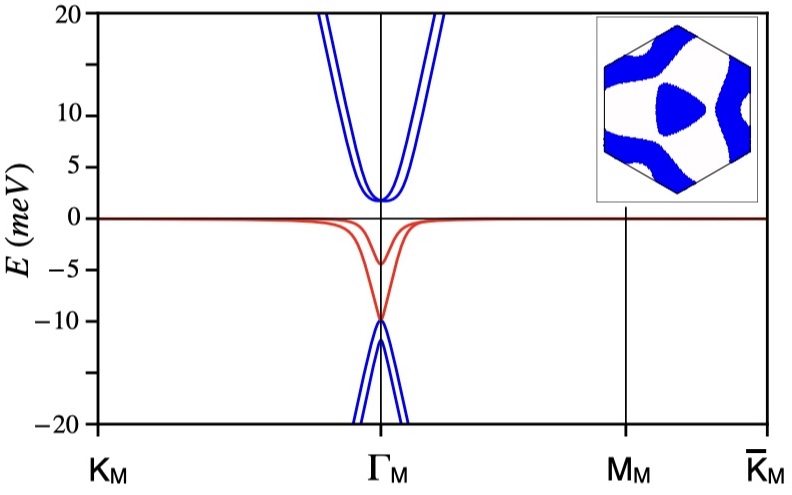}
	\caption{Many-body mean-field band-structure for $\nu = +2$. Parameters used: chemical potential  $\mu = 8.13\, meV$, $\xi = 8.18\, meV$, $\Ustar = 30\, meV$, $\kappa = 0.8$, with $\gamma_0 = 6.5 \, meV$ and $\gamma = 4.16\, meV$ at $T = 0.004\, meV$. Inset shows Brillouin zone with electron pocket relative to half-filling, shaded in blue.}
	\label{nu2bs}
\end{figure}

We conclude the mean-field results section by showcasing the quasiparticle bandstructures (Fig. \ref{halffillbs} - \ref{nu2bs}) for filling factors $\nu = \{0, +0.5,  +1,  +1.5, +2, \}$ at temperature $T = 0.004 \, meV \ll T_K^{(2)}$ below the Kondo coherence temperature.

The bandwidth of the mean-field quasiparticle dispersions (Fig. \ref{halffillbs} - \ref{nu2bs}) at $\Gamma_M$ remains at $2M$ but the bandwidth at $K_M$ is renormalized down, causing the bands to be incredibly flat away from the $\Gamma_M$ point. Using the renormalized mean-field hybridization $\gamma \approx 4 \, meV$ at low temperatures (Fig. \ref{gammawarmup}), we estimate the renormalized bandwidth at $K_M$ \eqref{MFbandwidth} to be
\begin{equation}
	\tilde{W}_{\text{MF}} = v^{\text{MF}}_D K_{\theta} \sim 0.01 M { \approx 0.03 \, meV \approx 0.4 \, K}.
\end{equation}
We find that the renormalized bandwidth $\tilde{W}_{\text{MF}}$ is approximately a tenth of the bare bandwidth \eqref{barebandwidth} and a good match to the Kondo coherence temperature $T_K^{(2)} \sim 0.01 \, meV$ (Fig. \ref{curieweiss}) in our mean-field theory numerics.


\section{Discussion}\label{sec:discussion}

In this paper, we have 
investigated the Song Bernevig model of twisted bilayer graphene,
taking advantage of the close analogy with heavy fermion and quantum
dot physics. 
We have developed
an auxiliary rotor
approach to describe the normal state across a  full range of
doping. Within the auxiliary rotor approach, we are able to capture
both the Kondo physics near integer filling factors and the strong
mixed valency near half integer filling factors.
One of the key points to arise from our study are 
discrepancies  between the ab-initio values of
the hybridization and onsite Coulomb interaction in the Song-Bernevig model, and
the corresponding values obtained from experiment. These discrepancies
suggest
renormalization processes, such as the effects of
phonons,  that may lie beyond the current model. Another key 
aspect of the physics is the presence of two scales in the problem -
a nominal Kondo temperature which establishes the topological
band-structure and a much lower scale at which the
flat bands lose their thermal entropy. 


Further experiments may help to 
test the foundation of the SB description. 
In conventional heavy fermion systems, the
presence of local moment behavior is immediately evident from the
Curie-Weiss behavior of the magnetic susceptibility 
\begin{equation}\label{}
\chi (T)\propto \frac{1}{T+\theta }
\end{equation}
To what extent can such Curie Weiss behavior be 
detected from a Maxwell-analysis of 
field dependent
compressibility measurements? It would 
be useful to use 
field-dependent compressibility
measurements to back-out the spin/valley susceptibility and directly
measure the size of the moment.  It would also be intriguing, in
future work, both theoretical and experimental, 
to observe and quantify  the 
charge redistribution between the
conduction and f-electrons as a function of gate voltage. 

Magnetic susceptibility measurements would be a key indicator of the predicted two Kondo temperatures ($T_K^{(1)}$ and $T_K^{(2)}$). A key issue is whether MATBG exhibits a Pauli paramagnetic region (indicating a heavy Fermi liquid) or whether Curie paramagnetism persists down to superconductivity or correlated insulator transition temperatures, as in the heavy fermion superconductors  NpPd$_5$Al$_2$ and CeCoIn$_5$\cite{Flint_NatPhys2008}.

The heavy fermion physics in MATBG raises an interesting possibility of parallels between STM tunneling  in MATBG and traditional heavy fermion systems. The specific manifestation of Kondo behavior will depend on the position-dependent tunneling amplitudes into the conduction states (dominating at the AB points) and the f-states (dominating at the AA points): the AB regions are expected to provide a broad feature characteristic of the Gamma-point light fermions, whereas the AA regions should image the the very narrow V-shaped density of states of the coherent f-states.  Generically, as in conventional heavy fermions, we expect a Fano feature corresponding to the interference between  conduction and f-state tunneling amplitudes\cite{maltseva:2009by}. The detailed form of Fano features, and their dependence on the incoherent scattering that will persist between  $T_K^{(1)}$ and $T_K^{(2)}$ are key issues for future investigation.

We note that our work does not yet include the effects of strain, particularly, "heterostrain", in which one layer is unixaxilly strained relative to the other.  Heterostrain  is known to preserve the Dirac points, while breaking the degeneracy between the renormalized Dirac cones in the same valley\cite{Kwan2021,jonahstrain2024}. Our two-temperature Kondo description is expected to remain valid in this situation because each Dirac cone will be associated with its own renormalized Dirac band-width, and the chemical potential will lie close to the $K_M$ or $\bar{K}_M$ Dirac cone, depending on whether the system is electron or hole doped. 
This increased Dirac bandwidth produced by heterostrain could however,  raise the lower Kondo temperature $T_K^{(2)}$ to the point where Fermi liquid behavior becomes apparent. Furthermore, the reduction of  nonlocal interactions relative to onsite terms by heterostrain \cite{jonahstrain2024} would provide additional support for the use of a  simplified treatment of local interactions model.

Although the work of Song and Bernevig emphasizes an ab-initio
description of the relevant interactions in MATBG, giving rise to 
an onsite Coulomb interaction in the range $60-100$meV, 
the measured $U\sim 30$meV \cite{cascade} is significantly smaller, suggesting that
there are important renormalization effects at work in the low-energy
physics of MATBG. A similar discrepancy is implicit in DMFT
approaches to MATBG\cite{haule_2019,bascones_2023}, which have encoded
the discrepancy in terms of a dielectric constant $\epsilon\sim 10$, a
factor of two or three larger than the accepted value for this
system. 

\begin{figure}[b]
	\centering
	\includegraphics[width = \columnwidth]{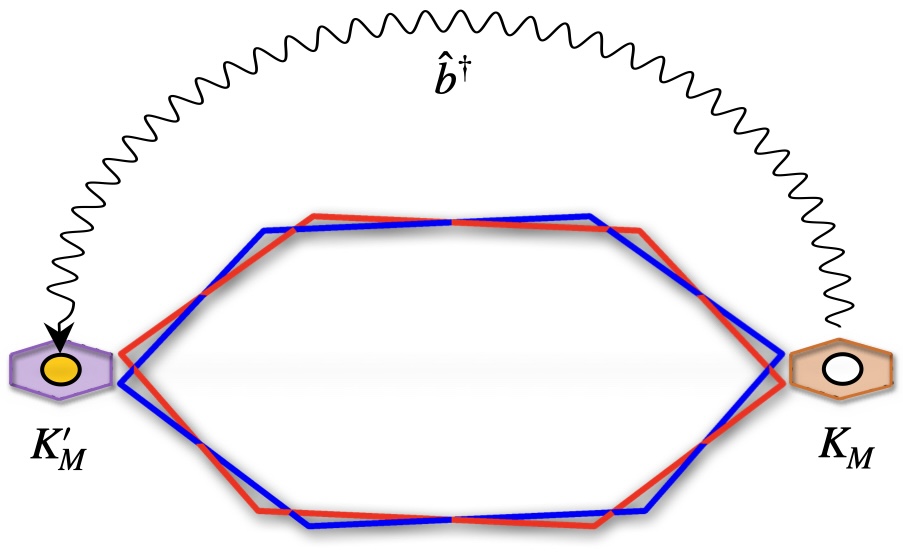}
	\caption{Schematic representation of the K-phonon process, whereby an f-electron in the valley $K_M$ is scattered to the other valley $K'_M$ through the absorption or emission of a K-phonon.}
	\label{Kphonon}
\end{figure}

A promising candidate for these renormalization
effects are the interactions of phonons with the valence fluctuations,
an effect which we now briefly discuss.  Recent $\mu$-ARPES experiments \cite{alietal2023}
have demonstrated a coupling of the flat f-bands with an inter-valley
optic phonon (Fig. \ref{Kphonon}) that gives rise to multiple satellites in the density of
states, separated by the optic K-phonon frequency of about $\omega_{0}\sim 150$meV.
These phonons modify the ``atomic'' Hamiltonian of the flat-band electrons
in MATBG, giving rise to a Holstein model of the form
\begin{equation}\label{eq:Holsteinsketch2}
H = \omega_{0}b\dg b +  g (b + b\dg )f\dg \tau_{x}f
\end{equation}
where $\tau_{x}$ denotes the valley spin operator and we have modelled
the optic phonon as a single Einstein mode. From this
Hamiltonian, we see that the valley spin of the f-electrons acts as a
force term on the optic phonons. Each time an f-electron is added to
the system, the phonons will valley polarize.  Moreover, since the
frequency of the Einstein mode is far greater than the characteristic
time-scales of valence fluctuations, this polarization will have time
to fully establish itself each time the valence fluctuates.   Such
polaronic effects are usually negligible in heavy fermion systems, where the
valence fluctuations are far faster than the
phonons \cite{Sherington76, Riseborough87}, have recently been observed to develop in 
conjunction with the slow charge fluctuations associated with 
strange-metal behavior in the 
heavy fermion compound YbAlB$_{4}$\cite{Kobayashi23}. 
Suppose we add an f-electron into a coherent superposition of valleys,
forming a state $\vert f^{1},\tau_{x}=1\rangle $, then the
resulting bosonic ground-state is a Glauber state given by 
\begin{equation}\label{}
\vert b\rangle  = e^{-|z|^{2}/2} e^{-z b\dg }\vert  0 \rangle .
\end{equation}
where $z = \frac{g}{\omega_{0}}\tau_{x}$.  In fact, we can replace
$z\rightarrow \sqrt{n_{b}}$, where $n_{b}$ is the number of bosons
that condense in response to the valence fluctuation. From this simple argument,
we see that the f-quasi-particle operator for this system must also
include the coherent state term of the bosons, i.e
\begin{equation}\label{}
f\dg_{qp} = e^{-n_{b}/2} e^{ -\sqrt{n_{b}} (f\dg  \tau_{x}f ) b\dg } f\dg
\end{equation}
where we have dropped the valley spin indices. The overlap between an
unrenormalized f-state and the quasiparticle is then $\sqrt{Z}=
\langle f_{qp}\vert f\dg  \vert 0 \rangle = 
 e^{-n_{b}/2}$.  This immediately implies that both the
hybridization and the repulsive $U$ will be renormalized as follows
\begin{equation}\label{eq:polaronrenorm}
\gamma_{0}\rightarrow \gamma^{*}_{0} = \sqrt{Z}\gamma_{0},\qquad
U\rightarrow U^{*} = Z^{2}U
\end{equation}
Experiment indicates that $U^{*}/U \sim 1/4$, implying that $Z\sim
1/2$ corresponding to $n_{b}= \ln 2\sim 0.7$ valley phonons associated
with each additional f-electron.

Our auxiliary rotor mean-field approach necessitated a reduction of the bare hybridization from $\gamma_0 = 25 \, meV$ to $\gamma_0 = 6.5 \, meV$- exceeding the expected polaron renormalization (Eq. \ref{eq:polaronrenorm})- to preserve the resetting behavior of the chemical potential $\mu[\nu]$ as a function of filling. We attribute this to the method's overestimation of the physical problem's relevant cutoff, yielding $D = 133 \, meV$ instead of $U/2 = 15 \, meV$. We leave this to be addressed in future studies.

We note following previous authors \cite{Fabrizio2019, Fabrizio2020, liu2023, shi2024,wang2024}, that in addition to the renormalization of the Coulomb
interaction, the valley phonon will introduce an attractive
Valley-Hunds interaction of the form $H_{ph}=- \frac{g^2}{\omega_{0}} (f\dg
\tau_{x}f)^{2}$. In fact, when we normal order this interaction, we
obtain 
\begin{equation}\label{eq:el-phint}
H_{ph} = -  \frac{g^2}{\omega_{0}} \left[ n_{f}+ 
:(f\dg \tau_{x}f)^{2}: \right]
\end{equation}
The first term in this expression is a local binding potential which
provides a natural contribution to the heavy fermion binding potential
we have introduced in this paper, on purely phenomenological grounds,
allowing us to tentatively identify
\begin{equation}\label{}
\kappa \Ustar  =  \frac{g^2}{\omega_{0}}
\end{equation}
Based on the model fitting the $\mu$-ARPES data \cite{alietal2023}, we obtain $\kappa \sim 0.6$, which is in the right ballpark for our mean field theory. However, estimating $\kappa$ through electron-phonon coupling strengths from literature poses a challenge for two reasons. First, the relevant coupling is between optical phonons and localized electron states at the AA-sites, whereas ab-initio results typically address Bloch states and acoustic phonons. Second, reported coupling strengths vary widely in the literature.

The coupling strength required to obtain $\kappa = 0.8$,  $g \approx 60 \, meV$ lies between literature extremes: an order of magnitude above some electron-phonon couplings \cite{phonon2}- which only consider phonon frequencies in an energy window smaller than the $150 \, meV$ optical K-phonon, yet an order of magnitude below Bloch electron-optical K-phonon couplings \cite{liu2023}. A recent study of optical K-phonons coupled to AA-localized states suggest lower coupling values but acknowledge significant parameter sensitivity \cite{shi2024}. It would be interesting, in future studies to use the $\mu$-ARPES satellite data \cite{alietal2023} to fit $g$ and predict $\kappa$.
Several authors have commented on the possibility for the second-term of Eq. \ref{eq:el-phint} to
play an important role in the superconductivity of MATBG, 
The full implications of this line of
reasoning will be explored in our future work. 


A second key aspect of our study is the presence of two Kondo
scales for the MATBG Kondo lattice - a large separation between the Kondo temperature $T_K^{(1)}$ associated with
the Withoff-Fradkin fixed point, which establishes the topology in the
flat bands, and a considerably lower coherence temperature
$T_K^{(2)}$, at which the flat bands entangle with the conduction sea
and lose their thermal entropy, forming a heavy Fermi Liquid (see Fig. \ref{scalingschem2}). This finding is consistent with an 
extensive temperature regime hosting a two-fluid system \cite{lightheavyliquid2021} comprising
coherent hybridized electrons at the $\Gamma$ point,
and thermal flat f-states around the $K_{M}$ points. Our mean-field
theory provides the following estimates for the two scales (see
\eqref{TK1X} and \eqref{TK2})
\begin{equation}\label{}
	T_{K}^{(1)}\sim \gamma, \qquad  T_{K}^{(2)} \sim { B}\left(\frac{T_{K}^{(1)}}{K_{\theta }v_{\star }} \right)^{2}  M 
\end{equation}
where ${ B = 2(1+a_{\star}^2 K_{\theta}^2) + \tilde \lambda^2(1-a_{\star}^2 K_{\theta}^2) \sim 10}$ is a dimensionless factor and our estimate of the lower Kondo temperature $T_K^{(2)}$ is the renormalized Dirac bandwidth around the moir\'e K-points \eqref{MFbandwidth}.

Although we can only crudely estimate  $T_{K}^{(1)}\sim \gamma
\sim { 50 - 100}$K, i.e $T_{K}^{(1)}/ (v_{\star }K_{\theta })\sim { 1/100}$,
so that $T_{K}^{(2)}\sim M/100 \sim 0.5 $ K, the
important point is that the high temperature Kondo cross-over will dominate
most of the measured temperature range. Indeed, it is likely
that true Fermi liquid behavior is interrupted by superconductivity or
insulating behavior, i.e the Kondo I fixed point dominates most of the
interesting temperature range, and that the Fermi liquid limit is
never actually attained. 

We will now discuss some of the implications of our of our model for electron transport properties. Below $T_K^{(1)}$, we expect the that as the temperature decreases due to the hybridization between the local moments and the c-electrons develops, the gap from the flat band edge to remote bands will open and widen. In a conventional Fermi liquid the inelastic electron scattering
rate is governed by the phase space for three-quasiparticle
excitations, which grows as the square of the energy $\omega^{2}$
\begin{equation}\label{}
\tau_{FL} (\omega)\propto \int_{0}^{\infty }
d\epsilon_{1}d\epsilon_2d\epsilon_{3} \delta (\omega - \epsilon_{1}-\epsilon_{2}-\epsilon_{3}
) \propto \omega^{2},
\end{equation}
which at finite temperature leads to a $T^{2}$ scattering rate. 
However, our theory indicates that in MATBG, 
that at temperatures that lie between the two widely separated
scales $T_{K}^{(1)}$ and $T_{K}^{(2)}$, the flat band-electrons are
thermalized. For these electrons, the main dissipation
process will involve scattering into an 
intermediate state containing a particle-hole pair 
of electrons near the $\Gamma$ point. 
The phase space for this process involves 
one intermediate f-state at  $\epsilon_{3}=0$ and two rapidly dispersing 
electrons near the $\Gamma$ point at energies $\epsilon_{1}$ and
$\epsilon_{2}$, with a corresponding 
phase space
\begin{equation}\label{}
	\tau_{NFL} (\omega)\propto \int_{0}^{\infty }
d\epsilon_{1}d\epsilon_2d\epsilon_{3} \delta (\omega- \epsilon_{1}-\epsilon_{2}-\epsilon_{3}
)\delta (\epsilon_{3}) \propto \omega,
\end{equation}
that is now linear in energy. When thermal factors are included, this
leads to a scattering rate that is linear in temperature and energy: a
marginal Fermi liquid\cite{fivefriends}.  
This
observation raises an intriguing prospect that the strong-coupling
fixed point created by Withoff-Fradkin scaling, with
thermal flat band f-states, might provide the essential phase space for the strange metal $T$-linear resistivity observed in MATBG.  This is an
interesting topic for future study. 
One of the other curious features of our model, is that the decay of a flat-band fermion into three light fermions will exhibit a $T^2$ scattering rate, which may explain the curious observation of at $T^2$ Hall angle \cite{lyu_strange_2021}– a feature that is absent in an electron-phonon model \cite{phononstrangemetal} for the strange metal behavior.

{\it Acknowledgments}

This work was supported by Office of Basic Energy Sciences, Material
 Sciences and Engineering Division, U.S. Department of Energy (DOE)
 under contract DE-FG02-99ER45790 (LLHL and PC). We gratefully
 acknowledge discussions with Eva Andrei, B. Andrei Bernevig,  Premi
 Chandra, Shahal Ilani, Daniel Kaplan, Xi  Dai, Eslaf Khalaf, 
  Alexei Tsvelik, Roser Valenti, Nikhil Tilak, and Zhenyuan Zhang.

\appendix
\begin{widetext}
\section{Coulomb Energy with Screening}\label{AppendixA}

Here we evaluate the Coulomb integral \eqref{screenyC}
\begin{equation}\label{A1.1}
U = \int_{\bx,\bx '} \rho (\bx )V (\bx -\bx ')\rho (\bx ') 
\end{equation}
where 
\begin{equation}\label{A1.2}
\rho (\bx ) = \frac{e^{-x^{2}/\lambda^{2}}}{\pi\lambda^{2}}
\end{equation}
is normalized electron density in the Wannier state, 
\begin{equation}\label{A1.3}
V (\bx -\bx ') =
\frac{e^{2}}{4\pi\epsilon\epsilon_{0}}
 \sum_{n}
\frac{(-1)^n}
{\sqrt{|\bx -\bx '|^{2}+ (2 d n)^{2} }}
\end{equation}
is the screened Coulomb interaction between electrons,
where $d$ is the distance to the back-gate(s). The summation 
$\sum_{n=0,1}$ for a single back-gate with a single image
charge per electron and $\sum_{n=-\infty}^\infty$
for a double back-gate, where there are 
the multiply reflected image charges of alternating sign. 

The result of the integral \eqref{result1} is
\begin{equation}\label{result1}
U = U_{0}F\left[\frac{d}{\lambda} \right]
\end{equation}
where 
\begin{equation}\label{result2}
U_{0}
= 
\sqrt{\frac{\pi}{2}}\frac{e^{2}}{4 \pi \epsilon_{0}\epsilon \lambda}
\end{equation}
is the unscreened Coulomb interaction and 
\begin{equation}\label{result3}
F[x]= \sum_{n} (-1)^{n}e^{2xn^{2}}{\rm Erfc}[\sqrt{2 x}n],
\end{equation}
describes the dependence of the screening on the distance to the
back-gate. For a single back-gate $\sum_{n}= \sum_{n=0,1}$, while for
a double back-gate, the sum is over all integers 
so that
\begin{equation}\label{}
F[x] =\left\{ 
\begin{array}{ll}
1 - e^{2x^{2}}{\rm Erfc}[\sqrt{2}x], & \hbox{single back-gate}.\cr
1 + 2 \sum_{n=1}^{\infty } (-1)^{n}e^{2x^{2} n^{2}}{\rm
Erfc}[\sqrt{2}xn], & \hbox{double back-gate}.
\end{array} \right.
\end{equation}

To obtain \eqref{result1}, we first 
substitute \eqref{A1.2} and \eqref{A1.3} into \eqref{A1.1} to obtain
\begin{equation}\label{}
U = 
\frac{e^{2}}{4 \pi \epsilon_{0}\epsilon}\int d^{2}xd^{2}x'
\frac{1}{(\pi\lambda^{2})^{2}}e^{- (x^{2}+x'^{2})/
\lambda^{2}}
 \sum_{n}
\frac{(-1)^n}
{\sqrt{|\bx -\bx '|^{2}+ (2 d n)^{2} }}.
\end{equation}
Changing variables to ${\bf R}= (\bx+\bx')/2$ and $\br =
\bx-\bx'$, so that $d^{2}xd^{2}x'=d^{2}rd^{2}R$,  we have 
\begin{equation}\label{}
U 
= 
\frac{e^{2}}{4 \pi \epsilon_{0}\epsilon}\int d^{2}rd^{2}R
\frac{1}{( \pi\lambda^{2})^{2}}e^{- (4R^{2}+r^{2})/2
\lambda^{2}}
 \sum_{n}
\frac{(-1)^n}
{\sqrt{r^{2}+ (2 d n)^{2} }}.
\end{equation}
Replacing $d^{2}r\rightarrow 2 \pi r dr$ and $d^{2}R\rightarrow 2\pi
RdR$, carrying out the first integral we obtain
\begin{eqnarray}\label{}
U &=& 
\frac{e^{2}}{4 \pi \epsilon_{0}\epsilon}
\overbrace {
\int_{0}^{\infty }
 \frac{2\pi R dR}{\pi\lambda^{2}}
e^{-2R^{2}/\lambda^{2}}}^{\frac{1}{2}}
\int_{0}^{\infty }\frac{2 \pi r dr}{\pi \lambda^{2}}
e^{- r^{2}/2
\lambda^{2}}
 \sum_{n}
\frac{(-1)^n}
{\sqrt{r^{2}+ (2 d n)^{2} }}\cr
&=& 
\frac{e^{2}}{4 \pi \epsilon_{0}\epsilon}
\int_{0}^{\infty }
\frac{ r dr}{\lambda^{2}}
e^{- r^{2}/2
\lambda^{2}}
 \sum_{n}
\frac{(-1)^n}
{\sqrt{r^{2}+ (2 d n)^{2} }}
\cr
&=& 
\frac{e^{2}}{4 \pi \epsilon_{0}\epsilon}
\int_{0}^{\infty }
 du
e^{- u}
 \sum_{n}
\frac{(-1)^n}
{\sqrt{2 \lambda^{2}u+ (2 d n)^{2} }}
= 
\frac{1}{\sqrt{2}}\frac{e^{2}}{4 \pi \epsilon_{0}\epsilon \lambda}
 \sum_{n} ( -1)^n
\int_{0}^{\infty }
 du
\frac{e^{- u}
}
{\sqrt{u+ a
n^{2} }}, 
\end{eqnarray}
where we have substituted $u = r^{2}/2\lambda^{2}$ and set $a =
2\frac{d^{2}}{\lambda^{2}}$.
Carrying out the final integral, 
\begin{equation}\label{}
\int_{0}^{\infty }
 du
\frac{e^{- u}
}
{\sqrt{u+ a
n^{2} }}
=
 \sqrt{\pi}e^{a n^{2}}{\rm Erfc}[\sqrt{a}n],
\end{equation}
we obtain the results in \eqref{result1}, \eqref{result2} and \eqref{result3}.

\section{Renormalized Anderson Model: Coulomb Blockade Physics and
Chemical Potential}\label{sec:phemchempot}

Here we present the detailed analysis of the variation in chemical potential
with filling factor as described in section \ref{coulombblockade}.
Let us consider the variation of the chemical potential $\mu$ as a function of filling factor $\nu$ in an atomic model,
	\begin{equation}\label{phenatomic}
		H_{A} (\bR ) = \frac{\Ustar}{2} (\hat  n_{\bR }-4)^{2} - \Ustar \kappa \nu \hat{n}_{\bR} - \mu \hat{n}_{\bR},
	\end{equation}
	where $-\Ustar \kappa \nu$ is a phenomenological heavy-fermion potential proportional to the backgate voltage. We rewrite the atomic Hamiltonian \eqref{phenatomic} as the following,
	\begin{equation}\label{eq:atomicmodel}
		H_{A} (\bR ) = \frac{\Ustar}{2} (\hat  n_{\bR }-4 - \kappa \nu)^{2} - \mu \hat{n}_{\bR},
	\end{equation}
and ignoring constant offsets to the Hamiltonian. The physics here is similar to a quantum dot. The stability of the quantum dot with $n = Q$ electrons requires that the ionization energies
	\begin{eqnarray}\label{phenomionization}
		\Delta E^{Q}_{\pm} = E_{Q\ \pm 1} - E_Q &=& \frac{\Ustar}{2} (Q\pm1-4 - \kappa \nu)^{2} - \mu (Q \pm 1) - \frac{\Ustar}{2} (Q-4 - \kappa \nu)^{2} + \mu Q \cr
		&=& \frac{\Ustar}{2} \pm \left( \Ustar (Q - 4) - \Ustar \kappa \nu - \mu\right) \cr
		&=& \frac{\Ustar}{2} \pm \left( \Ustar (1 - \kappa) \nu - \mu\right),
	\end{eqnarray}
to be positive, where in the last line we have used that the filling factor is defined as $\nu = Q - 4$. The ionization energies are positive provided that the chemical potential satisfies,
	\begin{equation}
		\frac{\Ustar}{2} > |\Ustar (1 - \kappa) \nu - \mu|.
	\end{equation}
	At integer filling factor $\nu$, the chemical potential jumps from $-\Ustar/2 + \Ustar(1-\kappa) \nu$ to  $\Ustar/2 + \Ustar(1-\kappa) \nu$ to compensate for the extra Coulombic cost. As the filling factor $\nu$ is tuned continuously from $\nu \rightarrow \nu + 1$, the emergent heavy-fermion potential compensates for $\Ustar \kappa$ of the Coulombic cost, hence the chemical potential drops by the same amount $\Ustar \kappa$ to maintain the filling of the localized state.
	We will now consider two extreme cases: the capacitive ($\kappa = 0$) and traditional heavy-fermion ($\kappa = 1$). For the capacitive scenario (Fig. \ref{fig:capacitormu}), there is no emergent heavy-fermion potential on the AA-sites, therefore the chemical potential has to bear the full Coulombic cost $\Delta \mu = \Ustar$ when the filling factor is increased by one $\Delta \nu = 1$.
\figwidth=0.8\textwidth
\begin{figure*}[tbh]\vspace*{-0cm}\centerline{\includegraphics[width=\figwidth]{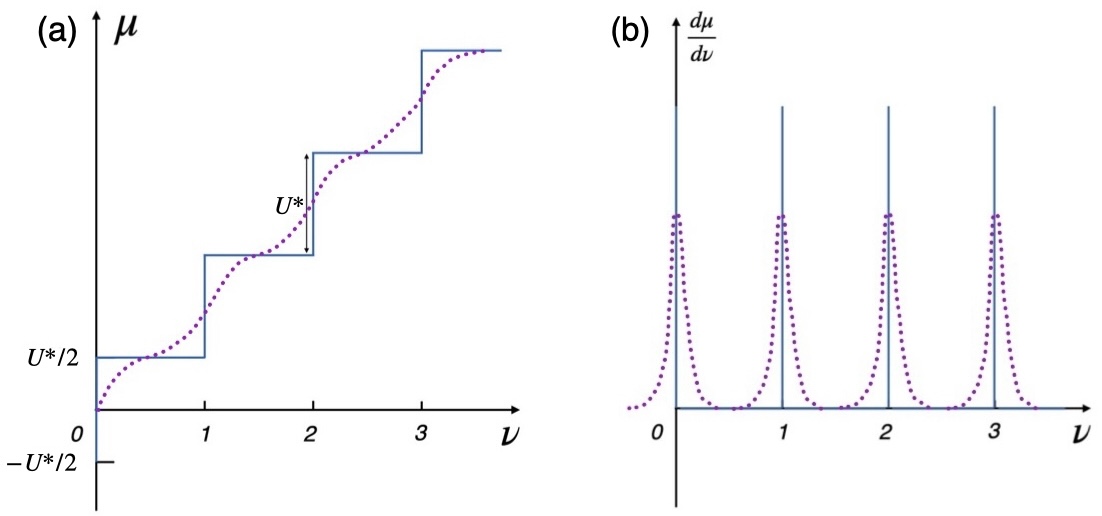}}\vskip
-0.2cm \caption{Sketch in blue of (a) the chemical potential $\mu$ and (b) the inverse compressibility $d \mu/ d \nu$ as functions of filling factor $\nu$ for the atomic limit of the renormalized Anderson model at zero temperature with $\kappa = 0$. There is no effective heavy-fermion potential and the chemical potential jumps by $\Ustar$ at each integer filling factor to compensate for the Coulombic cost of filling an extra atomic state. The inverse compressibility remains positive in this model. }\label{fig:capacitormu}\end{figure*}
	For a traditional heavy-fermion system ($\kappa = 1$, Figure \ref{fig:hfmu}), neutrality shifts due to the perfect compensation of the Coulombic costs by the heavy-fermion potential, provided by protons in the nucleus. Consequently, there is no shift in the chemical potential $\Delta \mu = 0$ when the filling factor is increased by one $\Delta \nu = 1$.
\figwidth=0.8\textwidth
\begin{figure*}[tbh]\vspace*{-0cm}\centerline{\includegraphics[width=\figwidth]{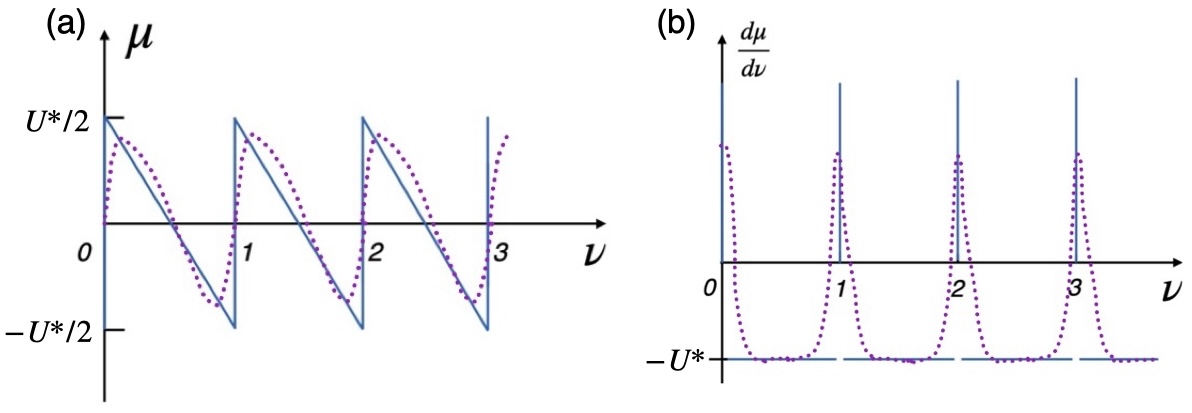}}\vskip
-0.2cm \caption{Sketch in blue of (a) the chemical potential $\mu$ and (b) the inverse compressibility $d \mu/ d \nu$ as functions of filling factor $\nu$ for the atomic limit of the renormalized Anderson model at zero temperature with $\kappa = 1$. The effective heavy-fermion potential perfectly compensates for the Coulombic cost for filling an extra atomic state, hence the chemical potential completely resets at each integer filling factor. The inverse compressibility can take negative values of order $-\Ustar$ in this model.}\label{fig:hfmu}\end{figure*}

\section{Derivation of the Flat Band Dirac Velocity in the Song-Bernevig Model} \label{sec:appendix}
\par Here we derive the  approximate expression \eqref{diracv} for the Dirac velocity
\begin{equation}\label{diracv2}
	{\rm v}_D \approx 
	3 \left(\frac{\gamma_K}{D}\right)^2 \frac{M}{D} \left(
	2(1+a_{\star}^2 K_{\theta}^2) + \tilde \lambda^2
(1-a_{\star}^2 K_{\theta}^2)\right){\rm v}_{\star }.
\end{equation}
at the $K_M$ points of the Song and Bernevig model, 
\begin{equation}\label{eq:startingpoint}
	H_{0}=\sum_{\substack{\bk' \in \text{MBZ}\\ \bG}}\Big[\sum_{a
a' \eta \sigma} \mathcal{H}_{a,a'}^{\left( \eta\right)} \left(\bk'-\bG \right)
c^{\dagger}_{\bk-\bG, a \eta \sigma} c_{\bk'-\bG, a' \eta \sigma}
 +
	\frac{\gamma_0}{\sqrt{N_s}}\sum_{\alpha a \eta \sigma} 
\big( 
	[\phi^{\left(\eta\right)}\left(\bk'-\bG, \gamma_0\right)]_{\alpha
a}
f^{\dagger}_{\bk' \alpha \eta \sigma
	}c_{\bk'-\bG,  a \eta  \sigma}  + \text{H.c.} \big)\Big].
\end{equation}
where the conduction electrons can take momenta outside the moir\' e Brillouin-zone and $\bG$ are reciprocal lattice vectors. Here

\begin{eqnarray}\label{eq:conductiondispersion}
	\scalebox{0.95}{$\mathcal{H}^{\left(\eta\right)}\left(\textbf{k}\right) = \begin{pmatrix}  & {\rm v}_{\star}\left(\eta k_x \alpha_0 + ik_y \alpha_z\right) \\
	{\rm v}_{\star}\left(\eta k_x \alpha_0 - ik_y \alpha_z\right) & M \alpha_x \end{pmatrix}$}.
\end{eqnarray}
The matrix form factor is
\begin{eqnarray}\label{eq:hybridizationmatrixr}
	\phi^{(\eta)}(\bk) \equiv \phi ^{\left(\eta\right)}\left(\textbf{k},\gamma_{0}\right) = e^{ \frac{-k^2 \lambda^2}{2} }\begin{pmatrix}
	 \alpha_{0} + a_{\star}\left(\eta k_x \alpha_x + k_y \alpha_y\right),  & 0_{2\times2} \end{pmatrix},
\end{eqnarray}
where $\gamma_{0}$  
and $a_{\star}$ set the magnitude and length scale of the
hybridization and $\lambda$ is a damping factor proportional to the real
space spread of the localized f-Wannier states. We also define the bandwidth $D = {\rm v}_{\star} K_{\theta}$, $\tilde{\lambda} = \lambda K_{\theta}$ which will be useful later.

We focus on the physics at the $K_M$ or $\bar{K}_M$ points and since $\phi_{\alpha a}^{\left(\eta\right)} \left(\bk\right)$ decays exponentially, we keep only the three MBZs surrounding the moir\' e $K$ point, preserving the $C_{3z}$ symmetry about that point (Fig. \ref{fig:vdtripod}).
\begin{figure}
	\centering
	\includegraphics[width = 0.4\columnwidth]{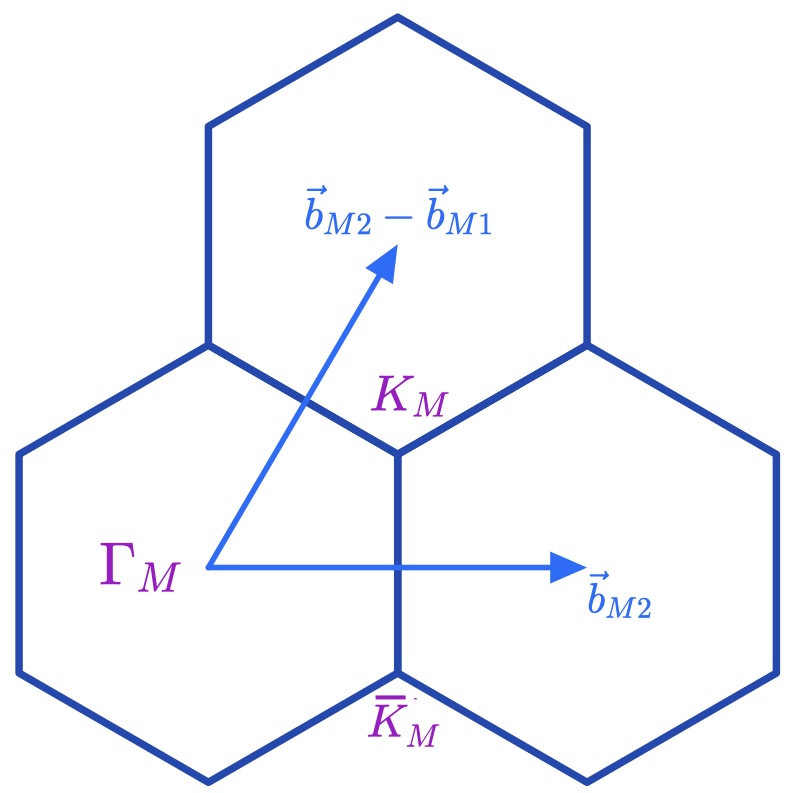}
	\caption{Sketch of the three moir\'e Brillouin zones around the $K_M$ point and the reciprocal lattice vectors.}
	\label{fig:vdtripod}
\end{figure}

Letting the eigenstates of the reduced model near the $K_M$ point to be
\begin{align}
    \footnotesize
    &|\Psi(\delta \bk)\rangle = \left(\sum_{\alpha}\psi_{\alpha}^{\left(f\right)}(\delta\bk)f^{\dagger}_{\bk\alpha,\eta \sigma} + \sum_{a}\big\{\psi_{a}^{\left(0c\right)}(\delta \bk)c^{\dagger}_{\bk a,\eta \sigma} + \psi_{a}^{\left(1c\right)}(\delta\bk)c^{\dagger}_{\bk-\bG_1, a,\eta \sigma} + \psi_{a}^{\left(2c\right)}(\delta\bk)c^{\dagger}_{\bk - \bG_2, a,\eta \sigma}\big\}\right)\biggr\rvert_{\bk=\textbf{K}_M + \delta\bk} |0\rangle \nonumber
\end{align}
we can rewrite the Schr\"odinger equation $\hat{H}_0|\Psi(\delta\bk)\rangle = E (\delta\bk) |\Psi(\delta \bk)\rangle$ in the first-quantized formalism  as
\begin{equation}
    H^{(\eta)}\left(\delta \bk\right) \Psi \left( \delta \bk \right) = E (\delta\bk) \Psi \left( \delta \bk \right),
\end{equation}
with the Hamiltonian matrix
\begin{equation}
	\footnotesize H_0^{(\eta)}\left(\bK_M + \delta \bk\right) = \begin{pmatrix} 0& \phi^{(\eta)}(\bK_M+\delta \bk)  & \phi^{(\eta)}(\bK_M - \textbf{b}_{M2} +\textbf{b}_{M1} +\delta \bk)  & \phi^{(\eta)}(\bK_M - \textbf{b}_{M2}  +\delta \bk) \\ \phi^{\dagger (\eta)}(\bK_M  +\delta \bk) & \mathcal{H}^{( \eta)}\left(\bK_M+\delta \bk \right) & 0 & 0 \\ \phi^{\dagger (\eta)}(\bK_M - \textbf{b}_{M2} +\textbf{b}_{M1} +\delta \bk)  & 0 & \mathcal{H}^{(\eta)}(\bK_M-\textbf{b}_{M2} +\textbf{b}_{M1} +\delta \bk) & 0 \\ \phi^{\dagger (\eta)}(\bK_M-\textbf{b}_{M2} + \delta \bk) & 0 & 0 & \mathcal{H}^{( \eta)}(\bK_M - \textbf{b}_{M2} + \delta \bk) \end{pmatrix}
\end{equation}
acting on the fourteen-dimensional spinor 
\begin{equation}
\Psi \left( \delta\bk \right) = \left( \psi^{(f)} \left( \delta\bk \right), \psi^{(0c)} \left( \delta\bk \right) , \psi^{(1c)} \left( \delta\bk \right), \psi^{(2c)} \left( \delta\bk \right) \right).
\end{equation}
Defining $\bG_0 = \textbf{0}$, $\bG_1 = \textbf{b}_{M2} - \textbf{b}_{M1}$ and $\bG_2 = \textbf{b}_{M2}$, the component form of the first-quantized formalism is
\begin{align}
	&\sum_{i=0,1,2} \phi^{(\eta)}\left(\bK_M - \bG_i + \delta \bk\right) \psi^{(ic)} = E \psi^{(f)} \label{eq:fschrodingerequation}\\
	& \phi^{\dagger (\eta)}\left(\bK_M - \bG_i+\delta \bk\right) \psi^{(f)} + \mathcal{H}^{(\eta)}\left(\bK_M - \bG_i + \delta \bk \right) \psi^{(ic)} = E \psi^{(ic)},\; i=0,1,2.
\end{align}
 Rearranging the second equation, we get $\psi^{(ic)} = \left(E - \mathcal{H}^{( \eta)}\left(\bK_M - \bG_i + \delta \bk\right)\right)^{-1} \phi^{\dagger (\eta)}\left(\bK_M - \bG_i + \delta \bk\right) \psi^{(f)}$ and inserting the expression into the first equation:
\begin{equation}\label{eq:felectronenergy}
	E \psi^{(f)} = \sum_{i=0,1,2} \phi^{(\eta)}\left(\bK_M - \bG_i + \delta \bk\right)\left(E - \mathcal{H}^{(\eta)}\left(\bK_M - \bG_i + \delta \bk\right)\right)^{-1} \phi^{\dagger(\eta)}\left(\bK_M - \bG_i + \delta \bk\right) \psi^{(f)}.
\end{equation}
The determinant of $\left(E - \mathcal{H}^{(\eta)}\left(\bk\right)\right)$ is 
\begin{eqnarray}\label{eq:det}
	E^4 - E^2(M^2 +2 {\rm v}_{\star}^2 k^2) +{\rm v}_{\star}^4 k^4
\end{eqnarray}
We neglect $\mathcal{O}\left(E^2\right)$ at $K_M$ and $\bar{K}_M$ and using the identity $\eta^2 = +1$,
\begin{align}
	\nonumber \scriptsize &\left(E - \mathcal{H}^{( \eta)}\left(\bk\right)\right)^{-1} \approx 
	\\ &\frac{1}{{\rm v}_{\star}^4|\bk|^4}\begin{pmatrix} (-M^2 -{\rm v}_{\star}^2 |\bk|^2)E \sigma_0 - M {\rm v}_{\star}^2 ((k_y^2 -k_x^2) \sigma_x + 2 k_x k_y \eta \sigma_y) & - {\rm v}_{\star}^3 |\bk|^2 (\eta k_x \sigma_0 + i k_y \sigma_z) + M {\rm v}_{\star}E (\eta k_x \sigma_x - k_y \sigma_y) \\
	- {\rm v}_{\star}^3 |\bk|^2 (\eta k_x \sigma_0 - i k_y \sigma_z) + M {\rm v}_{\star}E (\eta k_x \sigma_x - k_y \sigma_y) & -{\rm v}_{\star}^2|\bk|^2 E \sigma_0
	\end{pmatrix} \label{eq:inverseneglectesquared}.
\end{align}
Using Eq. \ref{eq:inverseneglectesquared} in Eq. \ref{eq:felectronenergy} and expand to first order in $\delta \bk$, where we also neglect the product of $\delta \bk$ and $E$, Eq. \ref{eq:felectronenergy} reduces to
\begin{eqnarray}
	E \psi^{(f)} &=& - \frac{3 \gamma_K^2 \left(M^2 + D^2\right) \left( K_{\theta}^2 a_{\star}^2 + 1  \right)}{ D^4} E \psi^{(f)}- \frac{3  M \gamma_K^2 \left(2\left(1 +a_{\star}^2 K_{\theta}^2\right) + \tilde \lambda^2 \left( 1- K_{\theta} a_{\star}\right) \left(K_{\theta} a_{\star} + 1\right) \right)}{K_{\theta} D^2}\left(\sigma_x \delta k_y - \sigma_y \delta k_x\right) \psi^{(f)} \cr \approx &-&\left( \frac{3 \gamma_K^2  \left( K_{\theta}^2 a_{\star}^2 + 1  \right)}{ D^2}E + \left(\frac{3M}{D}\right)\left(\frac{\gamma_K}{D}\right)^2 \left(2\left(1 +a_{\star}^2 K_{\theta}^2\right) + \tilde \lambda^2 \left( 1- K_{\theta} a_{\star}\right) \left(K_{\theta} a_{\star} + 1\right) \right){\rm v}_{\star}\left(\sigma_x \delta k_y - \sigma_y \delta k_x\right)\right) \psi^{(f)}
\end{eqnarray}
where the last line is obtained because $D \gg M$ and $\gamma_K = \gamma_0 e^{-\tilde \lambda^2 /2}$. We can identify a velocity $\tilde v_D$ and the Z factor from a self-energy treatment of the f-electrons, where the Dirac velocity at the $K_M$ points is 
\begin{equation}
v_D = Z \tilde v_D.
\end{equation}
The velocity
\begin{equation}
    \tilde v_D = 3 \left(\frac{M}{D}\right)\left(\frac{\gamma_K}{D}\right)^2 \left(2(1+a_{\star}^2 K_{\theta}^2) + \tilde \lambda^2 (1- K_{\theta} a_{\star})(K_{\theta}a_{\star}+1)\right) {\rm v}_{\star},
\end{equation}
and the Z factor is
\begin{equation}
    Z = \left(1 + 3\left(\frac{\gamma_K}{D}\right)^2  \left( K_{\theta}^2 a_{\star}^2 + 1  \right) \right) ^{-1} \approx 1
\end{equation}
because $(\gamma_K/D) \ll 1$. Rearranging yields the final result
\begin{equation}
	E \psi^{(f)} = -v_D \, \left(\sigma_x \delta k_y - \sigma_y \delta k_x\right)  \psi^{(f)},
 \end{equation}
 linearized near the $K_M$ point and 
 \begin{equation}
	v_D = 3 \left(\frac{M}{D}\right)
	\left(\frac{\gamma_K}{D}\right)^2 \left( 2(1+a_{\star}^2
	K_{\theta}^2) + \tilde \lambda^2 (1-
a_{\star}^{2}K_{\theta}^{2})\right) {\rm v}_{\star},
\end{equation}
corresponding to \eqref{diracv}. 
Taking the $a_{\star} = 0$ limit gives
\begin{equation}
   v_D \vert_{a_{\star}=0}= 3 \left(\frac{M}{D}\right) \left(\frac{\gamma_K}{D}\right)^2 \left( 2+ \tilde \lambda^2 \right) {\rm v}_{\star} 
\end{equation}
Repeating the calculation for the $\bar{K}_M$ point in the $\eta=+1$ valley results in $v_D \, \left(\sigma_x \delta k_y - \sigma_y \delta k_x\right) \psi^{(f)} = E \psi^{(f)}$, which has the same chirality as the Dirac cone at $K_M$ in the same valley. The Dirac cone structure of the topological heavy fermion model for the $K'_M$ and $\bar{K}'_M$ points in the $\eta=-1$ valley is $-v_D \, \left(\sigma_x \delta k_y + \sigma_y \delta k_x\right)$ and $v_D \, \left(\sigma_x \delta k_y + \sigma_y \delta k_x\right)$. Note that the chirality of the Dirac cones in the $\eta=-1$ valley is opposite to the Dirac cones in the $\eta=+1$ valley.
\section{Derivation of the Conduction Density of States in the
$\Gamma_{3}$ channel.}\label{sec:gamma3cdos}
Here we derive the conduction electron density of states\eqref{dos} in the
$\Gamma_{3}$ channel which hybridizes with the localized f-states. 
The conduction electron Hamiltonian has the form,
\begin{eqnarray}\label{eq:appconductiondispersion}
	\scalebox{0.95}{$\mathcal{H}^{\left(\eta\right)}\left(\textbf{k}\right)
	=
v_{\star }
(\eta k_{x}\tau_{1} \otimes \alpha_{0}-k_{y}\tau_{2} \otimes  \alpha_{z})
+\frac{1}{2}M  (1-\tau_{z})\otimes \alpha_{x} 
=
\begin{pmatrix}  & {\rm v}_{\star}\alpha_{x} (\vec{k}_{\eta }\cdot \vec{\alpha })\\
	{\rm v}_{\star}
(\vec{k}_{\eta }\cdot \vec{\alpha })\alpha_{x} 
 & M \alpha_x \end{pmatrix}$}.
\end{eqnarray}
Here the upper left two-by-two block 
is the $\Gamma_3$ irreducible
representation that hybridizes with the localized f-states, and the
lower two-by-two block is the $\Gamma_{1,2}$ irreducible
representation that do not hybridize.  
The Pauli matrices $\alpha_{\mu}\equiv (\alpha_{0},\vec{\alpha})$ ($\mu=0,3$) act on 
the two dimensional blocks, while the isospin matrices $(1,\vec{\tau})$
act on the two dimensional blocks that define the space of
representations, so that $P_{\Gamma_{3}}= \frac{1}{2}
(1+\tau_{z})$ projects into the $\Gamma_{3}$ channel. 
Here we have adopted the
notation $\vec{k}_{\eta }= (\eta k_{x},k_y)$, so that 
$\vec{k}_{\eta }\cdot \vec{\alpha }= \eta k_{x}\alpha_{x}+ k_y\alpha_{y}$.
The conduction sea is particle-hole symmetric
with energy eigenvalues  given by $\pm E_{\pm} (\bk )$, 
where 
\begin{eqnarray}
\label{eq:cdispersion}
	E_{\pm}(\bk)=  \pm \frac{M}{2} 
+ \sqrt{
\left(\frac{M}{2}\right)^2 +
\left(v_{\star}k\right)^2}
, 
\end{eqnarray}
and $k \equiv |\bk|$. The particle-hole symmetry allows us to write
the total conduction density of states as $\rho_{\rm TOT} (\omega)= \rho_{+}
(\omega)+\rho_{-} (\omega)$, where
\begin{eqnarray}\label{eq:rhoplusminusintermediate}
	\rho_{\pm}(\omega) &=& \sum_{\bk} \delta\left(\omega - |E_{\pm}(\bk)|\right) \cr &=& \frac{(2\pi)^2}{A_K}\int \frac{2 \pi k dk}{(2 \pi)^2} \delta\left(\omega - |E_{\pm}(\bk)|\right) \cr &=& \frac{\pi}{A_k} \int \frac{d k^2}{d E_{\pm}(\bk)} d E_{\pm} \delta\left(\omega - |E_{\pm} (\bk)|\right) \cr &=& \frac{\pi}{A_k} \left(\frac{d k^2}{dE_{\pm}}\Big\rvert_{E_{\pm} = |\omega|} \theta\left(|\omega|\mp M\right)\right),
\end{eqnarray}
where $A_k = 3 \sqrt{3}/2 K_{\theta}^2$ is the area of the hexagonal moir\'e Brillouin zone, and $\theta$ is the Heaviside step function. By noting $d k^2/dE_{\pm} = (2E_{\pm}(\bk) \mp M)/v_{\star}^2$, we get
\begin{equation}\label{eq:rhoplusminus}
	\rho_{\pm}(\omega) = \frac{2 \pi}{3\sqrt{3}} \frac{\left(2|\omega| \mp M\right)}{v_{\star}^2 K_{\theta}^2}\theta\left(|\omega|\mp M\right)
\end{equation}
as the c-electron density of states per spin per valley. Next we
project the conduction electron propagator into the $\Gamma_{3}$
channel by integrating out the $\Gamma_{1,2}$ electrons, 
\begin{equation}\label{eq:Gamma3propagator1}
	G_c^{\Gamma_3}\left(k, \omega\right) = 
P_{\Gamma_{3}}\frac{1}{(\omega - \mathcal{H}^{(\eta) } (\bk ))}P_{\Gamma_{3}}
=
	\left(\omega - v_{\star}^{2}
\alpha_{x} (\vec{k}_{\eta }\cdot \vec{ \alpha })
\frac{1}{\omega- M\alpha_{x}} (\vec{k}_{\eta }\cdot \vec{ \alpha })\alpha_{x}\right)^{{-1}}.
\end{equation}
Simplifying, we obtain
\begin{eqnarray}\label{eq:Gamma3xpropagator2}
	G_c^{\Gamma_3}\left(k, \omega\right) 
&=& (\omega^{2}-M^{2})
	\frac{\omega (\omega^{2}-M^{2}-(v_{\star} k)^{2})+M v_{\star} ^{2}[(k_{x}^{2}-k_{y}^{2})\alpha_{x}- ( 2 \eta k_{x}
k_{y})\alpha_{y}]}
	{[\omega (\omega^{2}-M^{2}-(v_{\star} k)^{2})]^{2}-
	(M v_{\star} ^{2}k^{2})^{2}}\cr &=& (\omega^{2}-M^{2})
	\frac{\omega (\omega^{2}-M^{2}-(v_{\star} k)^{2})+M v_{\star} ^{2}[(k_{x}^{2}-k_{y}^{2})\alpha_{x}- ( 2 \eta k_{x}
k_{y})\alpha_{y}]}
	{\left(\omega^2 - M^2\right) \left(\omega^2 - E_{+}(\bk)^2\right)\left(\omega^2 - E_{-}(\bk)^2 \right)}.
\end{eqnarray}
The  trace of Eq. \ref{eq:Gamma3xpropagator2} resolved along its poles gives,
\begin{eqnarray}\label{eq:trGamma3prop}
	{\rm Tr}G_{c}^{\Gamma_3} (k,\omega) &=&
	2\frac{\omega (\omega^{2}-M^{2} - (v_{\star} k)^{2})}
{(\omega^{2}-E_{+}^{2} (\bk))(\omega^{2}-E_{-}^{2} (\bk)) 
	} \cr &=&   \left\{ 
\frac{E_{-} (\bk)+M}{2E_{-} (\bk)+M} \left(\frac{1}{\omega - E_{-} (\bk)}+\frac{1}{\omega+ E_{-} (\bk)} \right)
\right. \cr
&+&\left. 
\frac{E_{+} (\bk)-M}{2E_{+} (\bk)-M} \left(\frac{1}{\omega - E_{+} (\bk)}+\frac{1}{\omega+ E_{+} (\bk)} \right)
 \right\}.
\end{eqnarray}
The conduction c-electron density of states, per spin per valley and
per orbital in the $\Gamma_{3}$ channel that hybridizes with the localized f-states is defined as the following,
\begin{eqnarray}\label{eq:Gamma3cdos1}
	\rho_c^{\Gamma_{3}}(\omega) &=& \frac{1}{2 \pi} \sum_{\bk} {\rm Im}{\rm Tr}G^{\Gamma_{3}}_{c} (k,\omega-i\delta ) = 
	\frac{1}{2}\sum_{\bk}\left\{ 
\frac{E_{-} (\bk)+M}{2E_{-} (\bk)+M}\delta (\omega - |E_{-} (\bk)|)
+ \frac{E_{+} (\bk)-M}{2E_{+} (\bk)-M}
 \delta ( |\omega| - |E_{+} (\bk)|)
	\right\} \cr &=& \frac{1}{2}\left\{ 
\frac{|\omega|+M}{2|\omega| +M}\rho_{-} (\omega)+ 
\frac{|\omega|-M}{2|\omega| -M} \rho_{+} (\omega)
 \right\},
\end{eqnarray}
where the factor of a half is to account for the valley
degeneracy. Recalling $\rho_{\pm} (\omega)$ from
Eq. \ref{eq:rhoplusminus}, our final expression for the conduction
c-electron density of states per spin per valley and per orbital in
the $\Gamma_{3}$ channel is
\begin{equation}\label{eq:Gammacdos2}
	\rho_c^{\Gamma_{3}}(\omega) =
	\frac{A}{D^2}\left(\frac{1}{2}
\left(|\omega| + M\right)
	\theta\left(M - |\omega|\right) +\frac{1}{2} |\omega|\,
	\theta\left(|\omega|-M\right)   \right) = \frac{A}{D^{2}}\times 
\left\{
\begin{array}{cc}
|E|
,& |E|>M\cr\cr
\frac{1}{2} (|E|+M),& |E|<M.
\end{array} \right.,
\end{equation}
where $A = 2 \pi /(3 \sqrt{3})$ and $D = v_{\star} K_{\theta}$. In the
main body of the text \eqref{dos} , we have dropped the 
$\Gamma_{3}$ superscript for clarity.
\section{Variance of the Hybridization}\label{sec:avghyb}

Here we evaluate the variance, or mean-square of the  hybridization
referred 
to in
\eqref{momint}. We recall the hybridization matrix \eqref{eq:hybridizationmatrix}, 
\begin{eqnarray}\label{eq:hybridizationmatrixq}
	\phi^{(\eta)}(\bk) \equiv \phi ^{\left(\eta\right)}\left(\textbf{k},\gamma_{0}\right) = e^{ \frac{-|\bk|^2 \lambda^2}{2} }\begin{pmatrix}
	 \alpha_{0} + a_{\star}\left(\eta k_x \alpha_x + k_y \alpha_y\right),  & 0_{2\times2} \end{pmatrix}.
\end{eqnarray}
The square of the hybridization matrix \eqref{eq:hybridizationmatrix}
is then
\begin{eqnarray}\label{hybsquared}
	\gamma_0^2\,\phi^{(\eta)}(\bk)\phi^{\dagger (\eta)}(\bk) = \gamma_0^2 e^{ -|\bk|^2 \lambda^2 }\Big(1+ a_{\star}^2 \left(k_x^2 + k_y^2\right) \alpha_0 - 2  a_{\star} \left( \eta k_x \alpha_x + k_y \alpha_y\right)\Big)
\end{eqnarray}
where the Pauli matrices
$\alpha_{\mu}\equiv (\alpha_{0},\vec{\alpha})$ ($\mu=0,3$) act on 
the two dimensional blocks.
The off diagonal terms of \eqref{hybsquared} average to zero when we take a momentum integrated average over a circle of radius $K_{\theta}$ because they are odd in $k_x$ and $k_y$. Averaging only the diagonal part of \eqref{hybsquared} over a circle of radius $K_{\theta}$ yields,
\begin{eqnarray}
	\overline{\gamma_0^2\,\phi^{(\eta)}(\bk)\phi^{\dagger (\eta)}(\bk)}&=& \frac{\gamma_0^2}{\pi K_{\theta}^2}\int_0^{2 \pi} \int_0^{K_{\theta}}  e^{ -k^2 \lambda^2 }\Big(1+ a_{\star}^2 k^2\Big) \alpha_0 \; dk\, d\theta \cr &=& \gamma_0^2 \Bigg[\frac{a_{\star}^2 + \lambda^2 - e^{-\tilde{\lambda}^2} \left(\lambda^2 + a_{\star}^2 \left(1+\tilde{\lambda}^2\right)\right)}{\lambda^2 \tilde{\lambda}^2}\Bigg] \alpha_0 \cr &\equiv& \overline{\gamma^{2}_0(k)} \alpha_0.
\end{eqnarray}
Using the approximate scales for the parameters in the SB model \cite{song21, Lau23} $\gamma_{0}=25$meV, $M=$3.7meV, $v_{\star }= -4.3$eV\AA, $K_{\theta
}=0.031 \hbox{\AA}^{-1}$, $a_{\star } = 65$\AA, and
$\lambda = 0.225 a_{M}= 29$\AA \ for the size of the Wannier states, 
which gives $\tilde{\lambda}=\lambda K_{\theta}= 0.90$, we calculate
the momentum integrated average of the hybridization over a circle of
radius $K_{\theta}$ to be
\begin{equation}
	\overline{\gamma^{2}_0(k)} = 1.891 \gamma_0^2 
\approx 2 \gamma_0^2 = 1250 \, ({\rm meV})^2,
\end{equation}
where the last equality follows from taking the bare hybridization $\gamma_0 = 25$meV.

\end{widetext}

\bibliography{Bigcombi}

\begin{thebibliography}{115}
\bibitem [{foo()}]{foot1}%
  \BibitemOpen
  \href@noop {} {\bibinfo  {journal} {{We note that one of the weaknesses of
  the auxilliary boson approach in its current form, is that the effective
  cut-off is given by the conduction band-width, about $2v_{\star}K_{\theta
  }=2D$ rather than $U$ as in \eqref{lowerTK} giving an artificialy enhanced
  $T_{K}^{(1)}$, which we we are adjusting for by reducing the effective Kondo
  coupling constant by adjusting the value of $\gamma_{0}$. A more realistic
  approach, to be explored in future work, will adopt a cut-off in the in the
  Matsubara sums $|\omega_{n}|<U$ associated with the electronic free energy in
  equations \eqref{eq:mfhalffillfreeenergy}, which we believe, will eliminate
  the need to make an adjustment in $\gamma_{0}$}}\ }\BibitemShut {NoStop}%
\end{thebibliography}

\end{document}